\documentclass[12pt,a4paper]{iopart}
\usepackage{iopams}
\usepackage{epsfig}
\usepackage{url}
\usepackage[dvips]{color}
\newcommand{\eqref}[1]{(\ref{#1})}
\newcommand{\zin}{z_{\rm in}}
\newcommand{\zout}{z_{\rm out}}

\definecolor{Black}{rgb}{.0,.0,.0}
\definecolor{Red}{rgb}{1.,0.,0.}

\begin{document}

\title{Looking the void in the eyes - the kSZ effect in LTB models}

\author{Juan Garc\'{\i}a-Bellido$^1$, Troels Haugb{\o}lle$^{1,2}$}
\address{$^1$ Instituto de F\'{\i}sica Te\'{o}rica UAM-CSIC,
Universidad Aut\'{o}noma de Madrid, Cantoblanco, 28049 Madrid, Spain,\\
$^2$ Department of Physics and Astronomy, University of Aarhus, DK-8000
Aarhus C, Denmark}
\ead{juan.garciabellido@uam.es, haugboel@phys.au.dk}

\begin{abstract}
As an alternative explanation of the dimming of distant supernovae it
has recently been advocated that we live in a special place in the
Universe near the centre of a large void described by a
Lema\^itre-Tolman-Bondi (LTB) metric. The Universe is no longer
homogeneous and isotropic and the apparent late time acceleration is
actually a consequence of spatial gradients in the metric. If we did
not live close to the centre of the void, we would have observed a
Cosmic Microwave Background (CMB) dipole much larger than that allowed
by observations. Hence, until now it has been argued, for the model to
be consistent with observations, that by coincidence we happen to live
very close to the centre of the void or we are moving towards it.
However, even if we are at the centre of the void, we can observe
distant galaxy clusters, which are off-centre. In their frame of
reference there should be a large CMB dipole, which manifests itself
observationally for us as a kinematic Sunyaev-Zeldovich (kSZ) effect.
kSZ observations give far stronger constraints on the LTB model
compared to other observational probes such as Type Ia Supernovae, the
CMB, and baryon acoustic oscillations. We show that current observations
of only 9 clusters with large error bars already rule out LTB models with
void sizes greater than $\sim 1.5$ Gpc and a significant underdensity,
and that near future kSZ surveys like the Atacama Cosmology Telescope
(ACT), South Pole Telescope (SPT), APEX telescope, or the Planck
satellite will be able to strongly rule out or confirm LTB models with
giga parsec sized voids. On the other hand, if the LTB model
is confirmed by observations, a kSZ survey gives a unique possibility of
directly reconstructing the expansion rate and underdensity profile of
the void.
\end{abstract}
\pacs{98.65.Dx, 98.80.Es, 98.80.-k\hspace{\stretch{1}} Preprint: IFT-UAM/CSIC-08-43}

\submitto{JCAP}
\maketitle

\section{Introduction}

Distant supernovae appear dimmer than expected in a matter-dominated
homogeneous and isotropic FRW universe. The currently favoured
explanation of this dimming is the late time acceleration of the
universe due to a mysterious energy component that acts like a
repulsive force. The nature of the so-called Dark Energy responsible
for the apparent acceleration is completely unknown.  Observations
seem to suggest that it is similar to Einstein's cosmological
constant, but there is inconclusive evidence. There has been a
tremendous effort in the last few years to try to pin down deviations
from a cosmological constant, e.g.~with deep galaxy catalogues like 
2dFGRS~\cite{2dFGRS} and SDSS~\cite{SDSS}, and extensive
supernovae surveys like ESSENCE~\cite{ESSENCE}, SNLS~\cite{SNLS}, 
and SDSS-SN~\cite{SDSS-SN}, and many more are planned for the near 
future e.g.~DES~\cite{DES}, PAU~\cite{PAU,Benitez2008}, BOSS~\cite{BOSS}
and JDEM~\cite{JDEM}.

In the meantime, our realisation that the universe around us is far
from homogeneous and isotropic has triggered the study of alternatives
to this mysterious energy.  Since the end of the nineties it has been suggested
by various groups~\cite{Hellaby:1998,Celerier:1999hp,Tomita:2000jj,Moffat:2005yx,
Garfinkle:2006sb,Enqvist:2007vb,Mattsson:2007tj,Wiltshire:2007zj,GBH:2008}
that an isotropic but inhomogeneous Lema\^itre-Tolman-Bondi 
universe could also induce an apparent dimming of the light of distant
supernovae, in this case due to local spatial gradients in the
expansion rate and matter density, rather than due to late
acceleration. It is just a matter of interpretation which mechanism is
responsible for the dimming of the light we receive from those
supernovae. Certainly the homogeneous and isotropic FRW model is more
appealing from a philosophical point of view, but so was the static
universe and we had to abandon it when the recession of galaxies was
discovered at the beginning of last century.

There is nothing wrong or inconsistent with the possibility that we
live close to the centre of a giga parsec scale void. Such a void may
indeed have been observed as the CMB cold
spot~\cite{Cruz:2006sv,Cruz:2006fy,Cruz:2008sb} and smaller voids
have been seen in the local galaxy distribution \cite{Frith:2003,Granett:2008}.
The size and depth of the distant observed voids, i.e.~$r_0 \sim 2$ Gpc
and $\Omega_M \sim 0.2$ within a flat Einstein-de Sitter universe,
seems to be consistent with that in which we may happen to
live~\cite{Tully:2007tp}, and could account for the supernovae
dimming, together with the observed baryon acoustic oscillations and
CMB acoustic peaks, the age of the universe, local rate of expansion,
etc.~\cite{GBH:2008}.

Moreover, according to the theory of eternal
inflation~\cite{Linde:1993xx}, rare fluctuations at the Planck
boundary may be responsible for the non-perturbative amplification of
local inhomogeneities in the metric, which would look like local voids
in the matter distribution~\cite{Linde:1994gy}. In the eternal
inflation approach one assumes to be a typical observer whose local
patch comes directly from a rare fluctuation within an inflationary
domain at the Planck scale. Moreover, since it is a rare fluctuation 
it should be highly spherically symmetric, and the theory predicts that 
we should live close to the centre of such a void~\cite{Linde:1994gy}. 
The size and depth of those voids depends on the theory of inflation on 
very large scales and may be a probe (perhaps the only probe) of the 
global structure of the universe.  

In fact, observations suggest that if there is such a large void, we should
live close to the centre, otherwise our anisotropic position in the void would 
be seen as a large dipole in the CMB.  Of course, we do observe a dipole, 
but it is normally assumed to be due to the combined gravitational pull of
the Virgo cluster, and the Shapley super cluster. There is always the
possibility that we live off-centre and we are moving towards the
centre of the void, so that the two effects are partially cancelled,
giving rise to the observed dipole. However, such a coincidence could
not happen for all galaxies in the void and, in general, clusters that
are off-centred should see, in their frame of reference, a large CMB
dipole. Such a dipole would manifest itself observationally for us as
an apparent kinematic Sunyaev-Zeldovich effect for the given cluster.

It is the purpose of this paper to study the very strong constraints
that present observations of the kSZ effect already put on the LTB
void models, and predict how near future observations from kSZ surveys
like ACT or SPT will strongly rule out (or confirm) LTB models with
giga parsec sized voids.

In section~2 we describe the general LTB void models, giving the corresponding
Einstein-Friedmann equations, as well as parameterisations of their
solutions. In a subsection we describe the GBH constrained model, where we
assume the Big Bang is homogenous, and thus the model depends on a single
function, the inhomogeneous matter ratio $\Omega_M(r)$. In section~3 we study 
the induced dipole for off-centred clusters and compute the size of the analogue
velocity of those clusters depending on the parameters of the void model. In 
section~4 we analyse present observations and give constraints on the model
from current observations. In section~5 we then explore the prospects that future 
experiments like ACT, SPT and Planck will provide for strongly constraining or
even ruling out LTB models of the universe. Finally, in section~6 we give our
conclusions.

\section{Lema\^itre-Tolman-Bondi void models}

The Lema\^itre-Tolman-Bondi model describes general radially symmetric
space-times and can be used as a toy model for describing voids in the universe.
The metric is
\begin{equation}\label{eq:metric}
ds^2 = - dt^2 + X^2(r,t)\,dr^2 + A^2(r,t)\,d\Omega^2\,,
\end{equation}
where $d\Omega^2 = d\theta^2 + \sin^2\theta d\phi^2$.
Assuming a spherically symmetric matter source with negligible pressure,
\begin{equation}
T^\mu_\nu = - \rho_M(r,t)\,\delta^\mu_0\,\delta^0_\nu\,,
\end{equation}
the $(0,r)$ component of the Einstein equations, $G^0_r = 0$, implies
$X(r,t)=A'(r,t)/\sqrt{1-k(r)}$, with an arbitrary function $k(r)$
playing the role of the spatial curvature parameter.
The other components of the Einstein equations read \cite{Enqvist:2006cg,
Enqvist:2007vb,GBH:2008}
\begin{eqnarray} \label{eq:FRW1}
{\dot A^2 + k\over A^2} + 2{\dot A\dot A'\over AA'} + {k'(r)\over
A A'} = 8\pi\,G\,\rho_M \,, \\
\dot A^2 + 2A\ddot A + k(r) = 0\,.
\end{eqnarray}
Integrating the last equation, we get
\begin{equation}
{\dot A^2\over A^2} = {F(r)\over A^3} - {k(r)\over A^2}\,,
\end{equation}
with another arbitrary function $F(r)$, playing the role of effective
matter content, which substituted into the first equation gives
\begin{equation}
{F'(r)\over A'A^2(r,t)} = 8\pi\,G\,\rho_M(r,t)\,.
\end{equation}
We can also use Eq.~\eqref{eq:FRW1} to define the critical density as
\begin{equation}
{\dot A^2 \over A^2} + 2{\dot A\dot A'\over AA'} = 8\pi\,G\,\rho_C(r,t)\,.
\end{equation}

The boundary condition functions $F(r)$ and $k(r)$ are specified by
the nature of the inhomogeneities through the local Hubble rate, the
local total energy density and the local spatial curvature,
\begin{eqnarray}
H(r,t) = {\dot A(r,t) \over A(r,t)} \,, \\
F(r) = H_0^2(r)\,\Omega_M(r)\,A_0^3(r)\,, \\[2mm]
k(r) = H_0^2(r)\Big(\Omega_M(r)-1\Big)\,A_0^2(r) \,,
\end{eqnarray}
where functions with subscripts $0$ correspond to present day values,
$A_0(r) = A(r,t_0)$ and $H_0(r) = H(r,t_0)$. With these definitions, the
$r$-dependent Hubble rate is written as \cite{Enqvist:2006cg,Enqvist:2007vb}
\begin{equation}\label{eq:hubblerate}
H^2(r,t) = H_0^2(r)\left[\Omega_M(r)\left({A_0(r)\over A(r,t)}\right)^3 +
(1-\Omega_M(r))\left({A_0(r)\over A(r,t)}\right)^2\right]\,,
\end{equation}
and we fix the gauge by setting $A_0(r)=r$.

For light travelling along radial null geodesics, $ds^2=d\Omega^2=0$, we have
\begin{equation}
{dt\over dr} = \mp {A'(r,t)\over\sqrt{1-k(r)}}
\end{equation}
which, together with the redshift equation,
\begin{equation}\label{eq:null}
{d\log(1+z)\over dr} = \pm {\dot A'(r,t)\over\sqrt{1-k(r)}}
\end{equation}
can be written as a parametric set of differential equations,
with $N=\log(1+z)$ being the effective number of e-folds
before the present time,
\begin{eqnarray}\label{eq:lightrays}
&&{dt\over dN} = - {A'(r,t)\over\dot A'(r,t)} \,,\\
&&{dr\over dN} = \pm {\sqrt{1-k(r)}\over\dot A'(r,t)}
\end{eqnarray}

\subsection{The constrained GBH model}

In general LTB models are uniquely specified by the two functions
$k(r)$ and $F(r)$ or equivalently by $H_0(r)$ and $\Omega_M(r)$, but
to test them against data we have to parameterise the functions, to
reduce the degrees of freedom to a discrete set of parameters.  For
simplicity in this paper we will use the constrained GBH model
\cite{GBH:2008} to describe the void profile. First of all, it uses a
minimum set of parameters to make a realistic void profile, and
secondly, it is assumed that the time to the Big Bang is constant for
spatial slices. The second condition gives a relation between $H_0(r)$
and $\Omega_M(r)$, and hence constrain the models to one free
function, and a proportionality constant describing the overall
expansion rate. Our chosen model is thus given by
\begin{eqnarray}
\hspace{-2cm}&&\Omega_M(r) = \Omega_{\rm out} + \Big(\Omega_{\rm in} - 
\Omega_{\rm out}\Big)
\left({1 - \tanh[(r - r_0)/2\Delta r]\over1 + \tanh[r_0/2\Delta r]}\right) \\
\hspace{-2cm}&&H_0(r) = H_0\left[{1\over \Omega_K(r)} -
{\Omega_M(r)\over\sqrt{\Omega_K^3(r)}}\ {\rm sinh}^{-1}
\sqrt{\Omega_K(r)\over\Omega_M(r)}\right] =
H_0 \sum_{n=0}^\infty {2[\Omega_K(r)]^n\over(2n+1)(2n+3)}\,,
\end{eqnarray}
where $\Omega_K(r) = 1 - \Omega_M(r)$, and the second equation follows
from the requirement of a constant time to a homogeneous Big Bang. We
use an ``inflationary prior'', and assume that space is asymptotically
flat, i.e.~in the following we set $\Omega_{\rm out}=1$. The model has
then only four free parameters: The overall expansion rate $H_0$, the
underdensity at the centre of the void $\Omega_{\rm in}$, the size of
the void $r_0$, and the transition width of the void profile $\Delta
r$. For more details on the model see Ref.~\cite{GBH:2008}.

\section{The CMB sky seen by off-centre observers}

\begin{figure}
\begin{center}
\includegraphics[width=0.65 \textwidth]{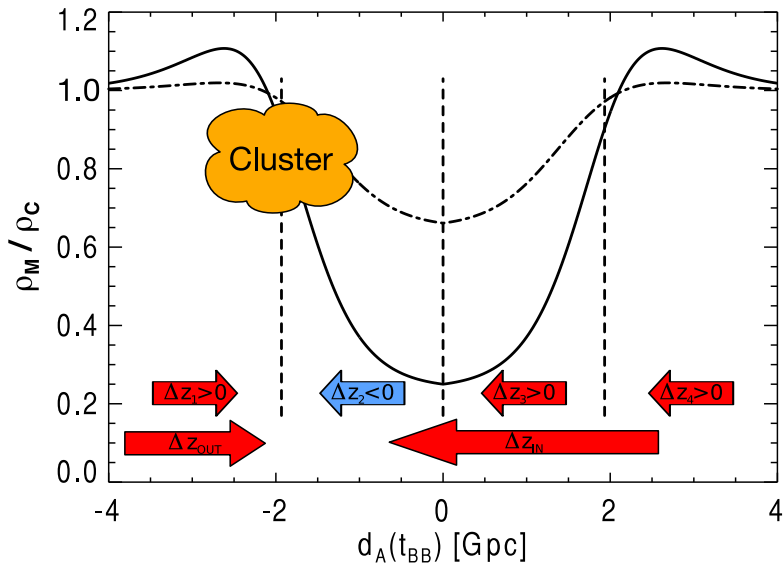}
\caption{An off-centre cluster of galaxies in a void will ``observe''
  CMB photons coming from the last scattering surface from all
  directions.  Due to the higher expansion rate inside the void,
  photons arriving through the centre (from the right in the figure)
  will have a larger redshift ($\Delta z_{\rm in}$), than photons
  arriving directly from the LSS (left, with $\Delta z_{\rm
    out}$). There is a subdominant effect due to the time-dependent
  density profile (the solid line corresponds to the current time,
  while the dot-dashed line to one tenth of the present time). With a
  larger underdensity at later times, we have $\Delta z_1 > \Delta
  z_4$, and $\Delta z_2 + \Delta z_3 < 0$, giving an overall
  difference $\Delta z_1 > \Delta z_2 + \Delta z_3 + \Delta z_4$ or,
  equivalently, a subdominant dipole with a blueshift towards the
  centre of the void. The overall effect is a blueshift away from the
  centre.}
\label{fig:setup}
\end{center}
\end{figure}

Imagine a cluster well embedded in a big void, but not exactly at the
centre, and consider photons reaching the cluster from the last
scattering surface (LSS). Because of the symmetry of the problem the
smallest and largest redshifts are found along the radial direction.
There are two effects contributing to the redshift of photons passing
through the void: The dominant effect is caused by the higher
expansion rate inside the void (see Eq.~\eqref{eq:lightrays}). The
photons coming from the farthest end of the void, crossing the centre,
are inside the void for the longest time, and thus have the biggest
redshift, while the photons arriving directly from the LSS along a
radial geodesic will be affected for the shortest amount of time, and
consequently suffer the least redshift. There is also a subdominant
effect due to the change in the gravitational potential or the matter
density: Photons coming from the farthest end of the void are first
gravitationally redshifted when entering the void, and then
gravitationally blueshifted after crossing the centre. On the other
hand, photons arriving directly from infinity are only redshifted.
There is a difference in the two redshifts, because of the time
dependence of the underdensity, and hence the gravitational
potential, with the subdominating effect leading to a small blueshift
towards the centre. This effect is in some sense a large scale
Rees-Sciama effect~\cite{RS:1968}, see Fig.~\ref{fig:setup}.

Consequently, in the ideal case of a spherical void, and a well
embedded cluster, the cluster observer will see an almost perfect
dipole in the CMB, aligned along the radial direction, and with the
blueshift pointing away from the centre of the void, where the
observer is (see Fig.~\ref{fig:setup}). The detailed effect of a
spherical void on the CMB sky of an off-centre observer has been
calculated in \cite{Alnes:2006pf}, and it is shown that for small
distances from the centre of the void the dominating term is a dipole.
For simplicity, in this paper we will estimate the change on the CMB
sky as a pure dipole.  To find the amplitude it is then enough to
integrate two radial light rays one going towards the centre, and the
other away from the centre of the void. The calculation of the grid of
parameters is done using our public {\tt easyLTB} program
\cite{easyLTB}, but we have also used the completely different
approach of Taylor expansion around an Einstein de Sitter solution, as
detailed in \cite{GBH:2008}, and checked that we get the same result
for the parameters in Fig.~\ref{fig:dipoles}.  It should be noted that
this dipole approximation breaks down when the effective size of the
void on the sky, as observed from the cluster, becomes too small
(i.e.~less than $\sim 2\pi$), or possibly when the density or Hubble
expansion profiles have very contrived time dependence.

Motivated by the observed size of our CMB dipole, we assume that we
are located at the centre of the void, and observe clusters in the
light cone at different redshifts.  Each cluster is then observed at a
certain time $t_{cl}(z)$, related to its redshift. In order to find
the dipole as seen by the cluster, we integrate
Eq.~\eqref{eq:lightrays} along the radial axis in the positive and
negative direction, and always backwards in time, with a starting time
$t_{cl}(z)$, and an ending time $t_{LSS}\sim 10^5$ yr. The size of the
dipole can be easily calculated from the temperature seen by observers
in the cluster in different directions, $T(\theta) =
T_*/(1+z(\theta))$, where $T_*$ is the temperature of the LSS. If we
now look in opposite directions, towards and away from the centre of
the void, we find
\begin{equation}
\frac{\Delta T}{T}_{\rm dipole} = {T(\theta) - \hat T\over\hat T} = 
{|T_{\rm in} - T_{\rm out}|\over T_{\rm in} + T_{\rm out}} = 
\frac{|\zin - \zout|}{2 + \zin + \zout}\,,
\end{equation}
where $\hat T=(T_{\rm in} + T_{\rm out})/2$ is the mean temperature
observed at the location of the cluster, and $z_{\rm in/out}$ are the
redshifts to the LSS for radially ingoing/outgoing light rays.

\subsection{The kinematic Sunyaev-Zeldovich effect}

The hot gas (mainly the electrons) inside a cluster inverse Compton
scatters CMB photons, changing their frequency distribution. The
thermal Sunyaev-Zeldovich effect~\cite{Sunyaev:1970,Sunyaev:1972} is
the main effect, redistributing low energy photons to higher energies
due to upscattering by the hot cluster gas, but also the intra-cluster
gas works as a mirror rescattering photons from all directions towards
the observer. If the CMB sky observed by the cluster is different than
the CMB sky observed by us, there will be an additional change
in the spectrum, which is known as the kinematic Sunyaev-Zeldovich
effect \cite{Sunyaev:1980nv}. Because of rotational symmetry
around the axis associated with the line of sight, only changes
projected along this axis will have an observational impact.  In the
LTB model, because the observer is supposed to be at the centre, an
off-centred cluster will observe a CMB dipole exactly along the line
of sight, which to first order in the perturbation is
indistinguishable from a kinematic dipole due to the peculiar velocity
of the cluster.  The change in the photon intensity is given as
\cite{Sunyaev:1980nv,Philips:1995,Birkinshaw:1999}
\begin{equation}\label{eq:kSZ}
\frac{\Delta I_\nu}{I_\nu} = - \frac{x e^x}{e^x-1}
  \int \sigma_T n_e {\boldsymbol{v}_p \over c} \cdot d\boldsymbol{l}\,,
\end{equation}
and thus
\begin{equation}\label{eq:TkSZ}
\frac{\Delta T_{\rm kSZ}}{T_{\rm rad}} = - \frac{x^2 e^x}{(e^x-1)^2}
  \int \sigma_T n_e {\boldsymbol{v}_p \over c} \cdot d\boldsymbol{l}
\simeq - \beta_p\,\tau_e\,\frac{x^2 e^x}{(e^x-1)^2}\,,
\end{equation}
where $x=h\nu / kT$, $\tau_e$ is the cluster optical depth, and the
apparent peculiar velocity is related to the temperature dipole as
\begin{equation}
\beta_p = \frac{v_p}{c} = \frac{\Delta T}{T}_{\rm dipole}\,.
\end{equation}

\begin{figure}
\begin{center}
\includegraphics[width=0.32 \textwidth]{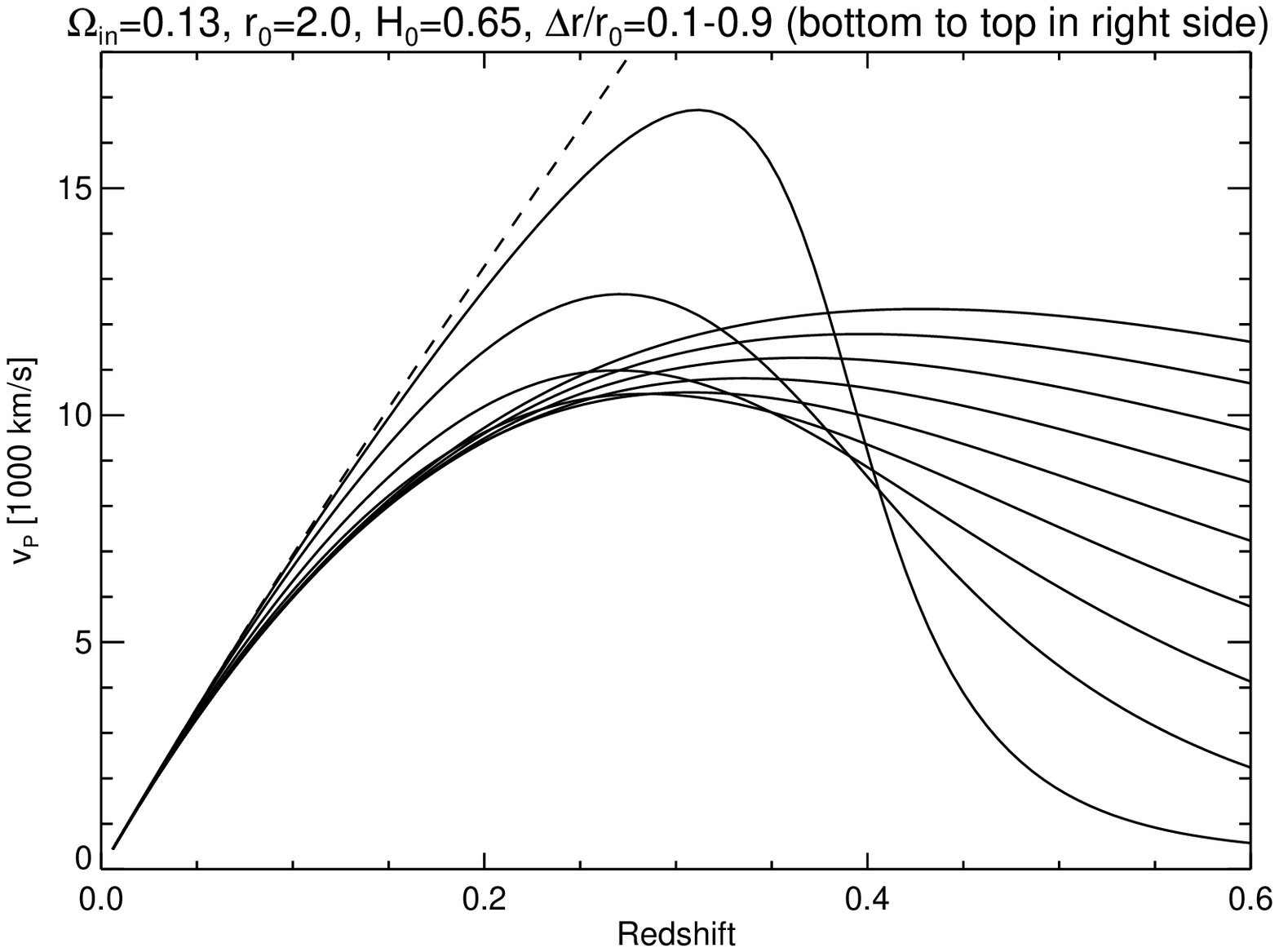}
\includegraphics[width=0.32 \textwidth]{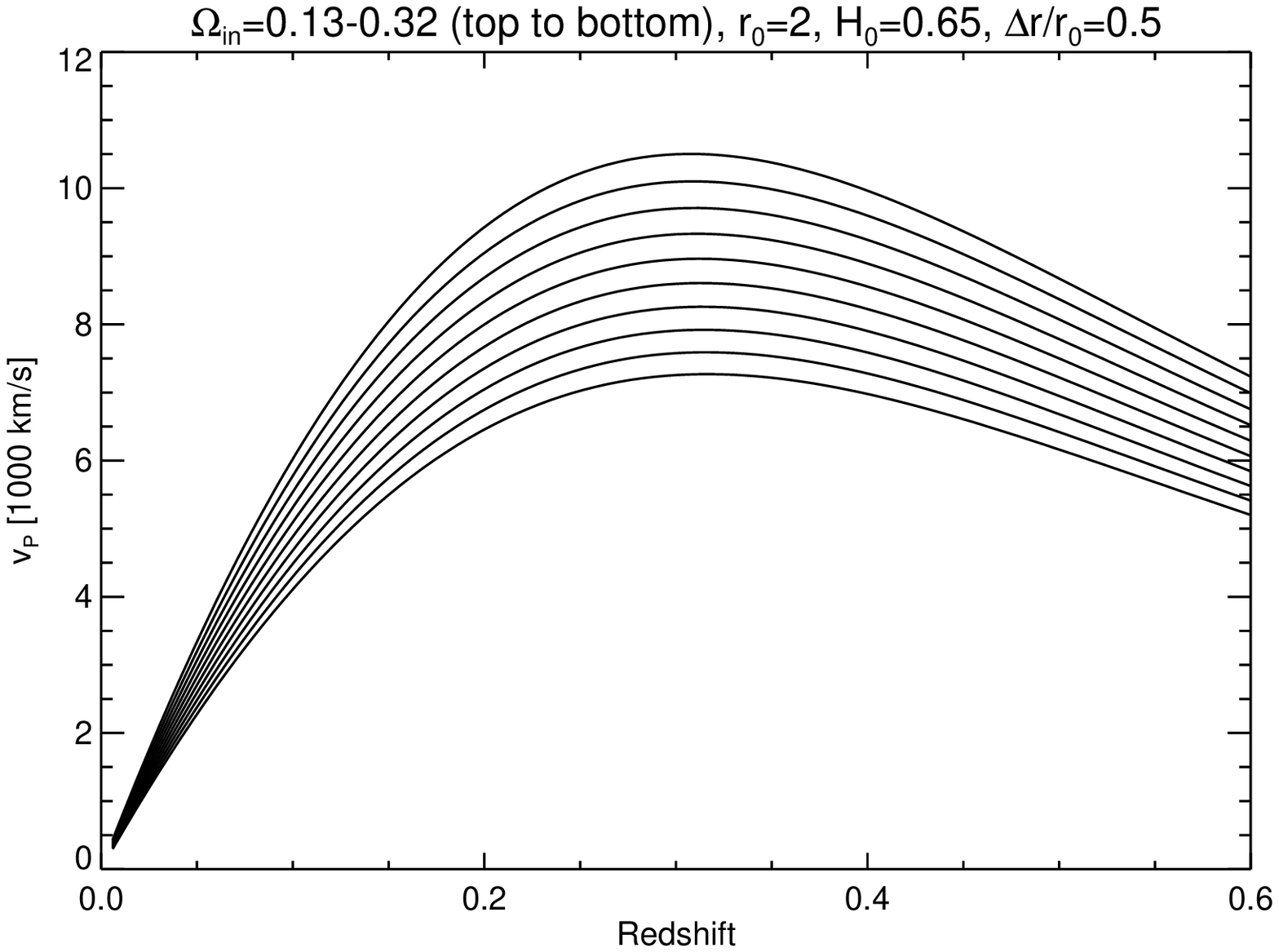}
\includegraphics[width=0.32 \textwidth]{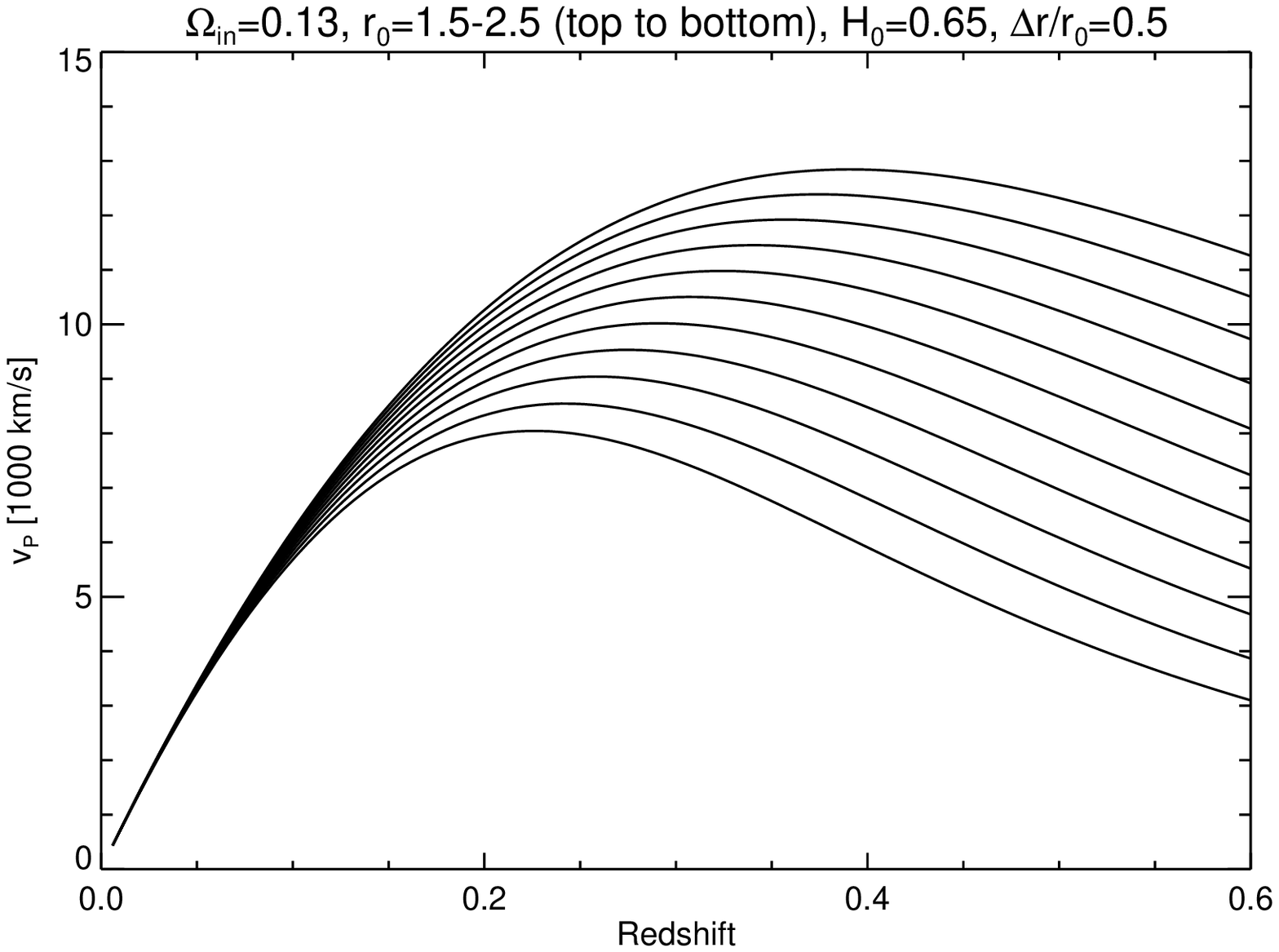}
\caption{Examples of the size of the dipole for different parameters
of the constrained GBH model \cite{GBH:2008}. The dashed line in the
left figure is the first order approximation given in
\cite{Alnes:2006pf}.}\label{fig:dipoles}
\end{center}
\end{figure}

\noindent
Even though the kSZ signal from a single cluster can be interpreted as
a peculiar velocity, a large void would result in a systematic trend
for all clusters (with a certain scatter given by the intrinsic
peculiar velocities of the clusters), with an average redshift or positive
apparent velocity (see Fig.~\ref{fig:dipoles}).  The average
velocity as a function of the distance would give a direct and unique
handle on the density and velocity profile of the void. At the same
time, the absence of such a systematic average peculiar velocity in
future kSZ surveys can strongly rule out a void of any appreciable
size.
 
\section{Constraints from current observations}
\begin{figure}
\begin{center}
\includegraphics[width=0.45 \textwidth]{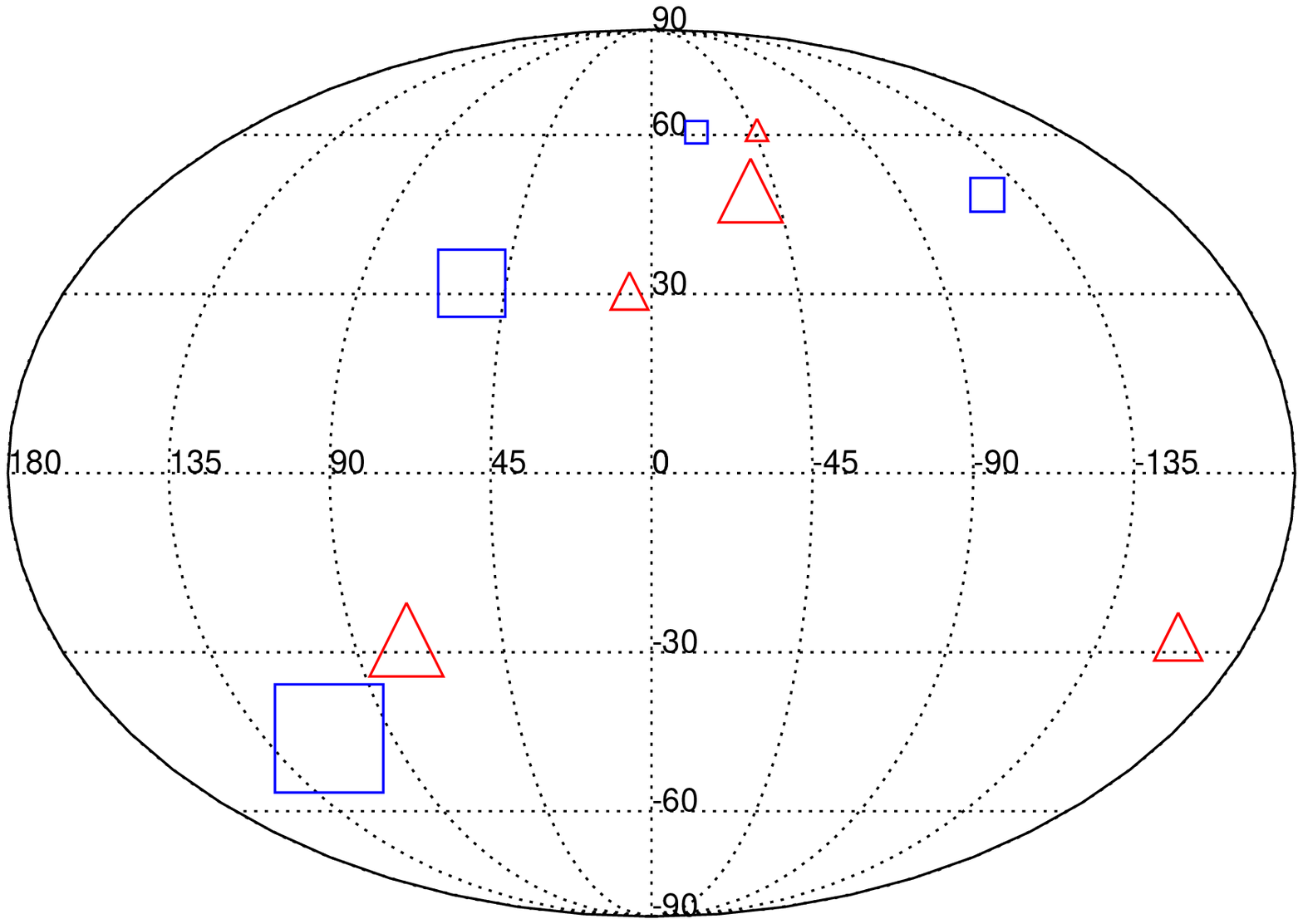}
\includegraphics[width=0.45 \textwidth]{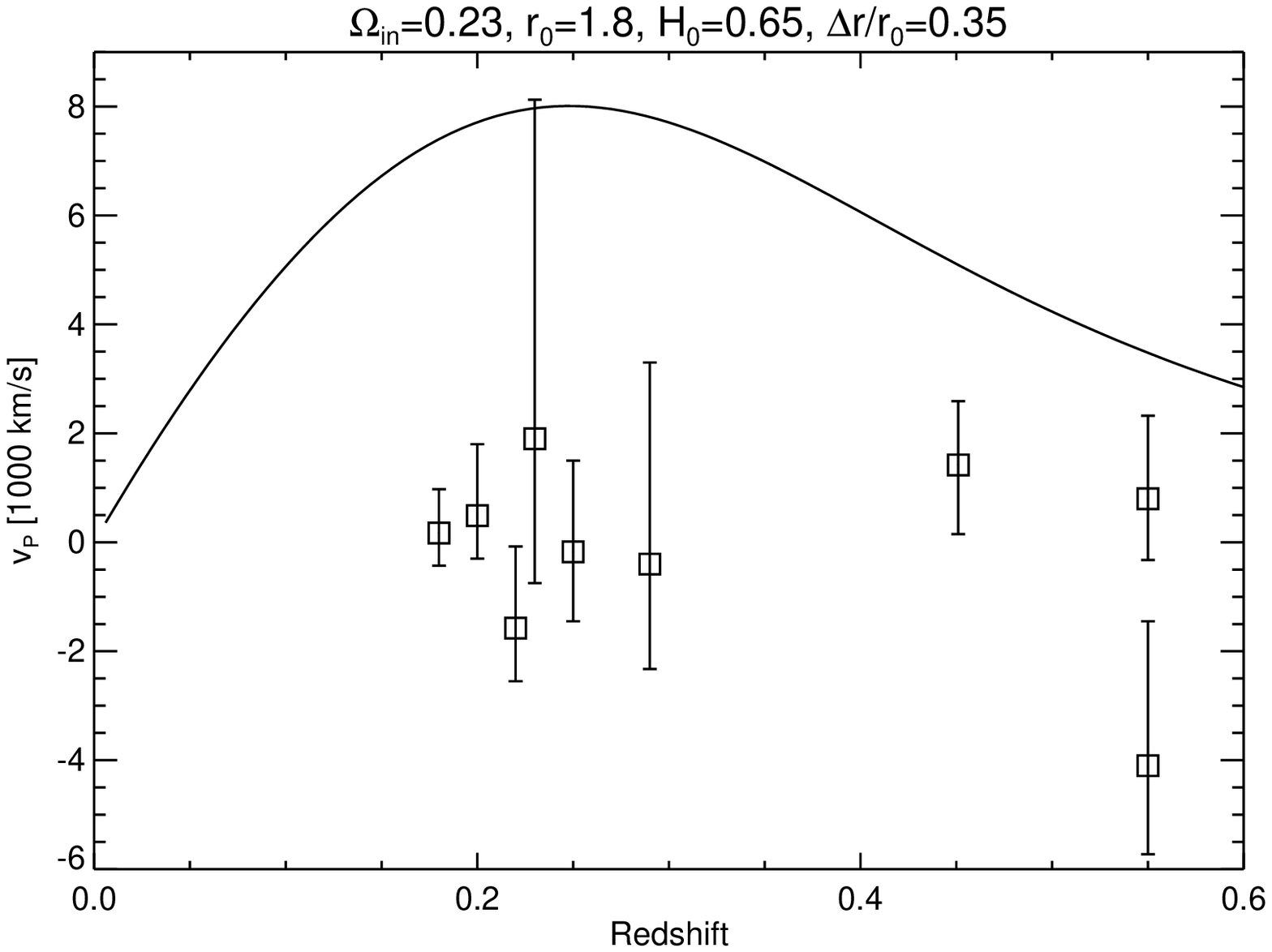}
\caption{The angular and redshift distribution of current observations together
with the predicted dipole distribution for a giga parsec sized void model. Red triangles
and blue squares represent positive and negative peculiar velocities respectively, with the
size of the symbol indicating the magnitude of the velocities.}\label{fig:obs}
\end{center}
\end{figure}

We have compiled a set of 9 clusters with kSZ measurements from
\cite{Hopzafel:1997,Benson:2003,Kitayama:2004} (see table \ref{tab:ksz}
and Fig.~\ref{fig:obs}).
Currently both systematic and random errors are very large, but even though
observations are scarce they already give very interesting bounds on the size of a
possible void.

We have used the observations to compare with the constrained GBH
model, making a grid of models with different parameters (see table
\ref{tab:priors}), and calculated the likelihood of each model. From the
literature we only have the (asymmetric) 68\% confidence limits for
the observations, and not the full probability distributions. To
respect the asymmetric error bars, while keeping the probability
distribution close to Gaussian, we use the $\epsilon$-skew-normal
($\epsilon$SN) distribution \cite{epsilonskew:2000}, which has the
probability density function (PDF)
\begin{equation}
f(v;v_i,\sigma^-,\sigma^+) = \frac{1}{\sqrt{2\pi}\left[\sigma^- + \sigma^+\right]}
  \exp \left [- \frac{\left[v - v_i\right]^2}{2\,\,\sigma^{\pm\,2}} \right] \,,
\end{equation}
where $\sigma^\pm=\sigma^-$ when $v\le v_i$, and $\sigma^\pm=\sigma^+$
when $v>v_i$, to model the individual data points.  The $\epsilon$SN
model is a reasonable model for the data, if we do not have knowledge
about the detailed likelihood for each data point, since the
probability distribution is continuous up to the first derivative, the
maximum likelihood is at $v_i$, and 68\% of the PDF is in the interval
[$v_i-\sigma^-,v_i+\sigma^+$], as required by the data. The mean value
is not at $v_i$ reflecting the fact that the distribution is skewed.

When comparing our models to the observations we not only have to take
into account measurement errors as given in the literature but also
the fact that, if the model is correct, the apparent peculiar
velocities of the clusters are the sums of the apparent dipole
velocities from the void, and the intrinsic peculiar velocities. In
the following we assume that the clusters, due to their high redshift,
have uncorrelated intrinsic velocities, though for future observations
of small fields with many clusters this may not be true (see
e.g. \cite{Haugboelle:2006,Hui:2005,Monteagudo:2005}). The peculiar
velocities are assumed to be normal distributed, and are added in
squares to the measurement errors.

For the full sample (see table \ref{tab:sample} for sub samples) the
average cluster velocity, excluding any systematic shift, and
correctly weighted\footnote{We use the $\epsilon$SN distribution for
  each data point, and weight the data according to $w_i = 1 /
  \sigma^2$, where the average standard deviation is the harmonic mean
  of the error bars: $2 / \sigma = 1 / \sigma^- + 1 / \sigma^+$.
  Using these weights the PDF of the average velocity is found as
  $f_{\bar v}(v) = \sum_i w_i f_i(v) / \sum_i w_i$, while the standard
  deviation is found by Monte-Carlo sampling of the data
  distributions: $\sigma_v = < \sum_i w_i (v_i - \bar v)^2 / \sum_i
  w_i >$, with $v_i$ drawn on random from $f_i(v)$.}
is $\bar v = 320^{+440}_{-400}$ km s$^{-1}$, while the
standard deviation among the cluster velocities is $\sigma_v =
1630^{+400}_{-350}$ km s$^{-1}$. This is a fairly large velocity scatter;
much larger than expected from linear theory. Given that the clusters in
current kSZ surveys are among the most massive clusters known, we
would expect radial peculiar velocities on the order of $\sim 400$ km
s$^{-1}$ from linear theory \cite{Monteagudo:2005}, which is a factor
of 4 lower than what is found from the internal scatter in our data set.

To estimate the typical peculiar velocity scatter $\sigma_{\rm pv}$ we
will use two different values: The average velocity as expected in a
$\Lambda$CDM model, fixed at 400 km s$^{-1}$, and the scatter of the
current data set, fixed at 1600 km s$^{-1}$. We could include
evolution in the velocities $-$ according to linear theory they grow
with the growth factor $-$ but since current data have very large
errors, it would not make much of a difference.

Summarising, the total log-likelihood for a given model is
\begin{equation}\label{eq:lobs}
-2 \ln \mathcal{L} = \sum_{i} \left[\frac{v_i - \alpha v(\vec p,z_i) 
                       + v_{\rm sys}}{[\sigma^{\pm\,2}_i + \sigma^2_{\rm pv}]^{1/2}}\right]^2\,,
\end{equation}
where $v_i$ is the observed value of the velocity, $v(\vec p, z_i)$ is
the value according to the model, with parameters $\vec p$, and
$\sigma^\pm$ is $\sigma^+$ if $v > v_i$ and else $\sigma^-$. According
to \cite{Benson:2003} there can be significant systematic errors in
the data, and we allow for this by adding a systematic shift in the
values: $v_{\rm sys}$. Below we use the systematic shifts $v_{\rm
  sys}=0$, and $v_{\rm sys}=750$ km s$^{-1}$.  
The size of the dipole is slightly overestimated
by our model. First of all because the effect of the void on the CMB
sky is not a pure dipole, and second because at large distances the
projected size of the void on the sky of the cluster becomes less than
$2\pi$, and then the dipole part is even less. To model it we include
an empirical rescaling factor $0<\alpha\le1$, and to estimate the effect
error we use $\alpha=1$ and $\alpha=0.8$.

\begin{figure}[!ht]
\begin{center}
\includegraphics[width=0.45 \textwidth]{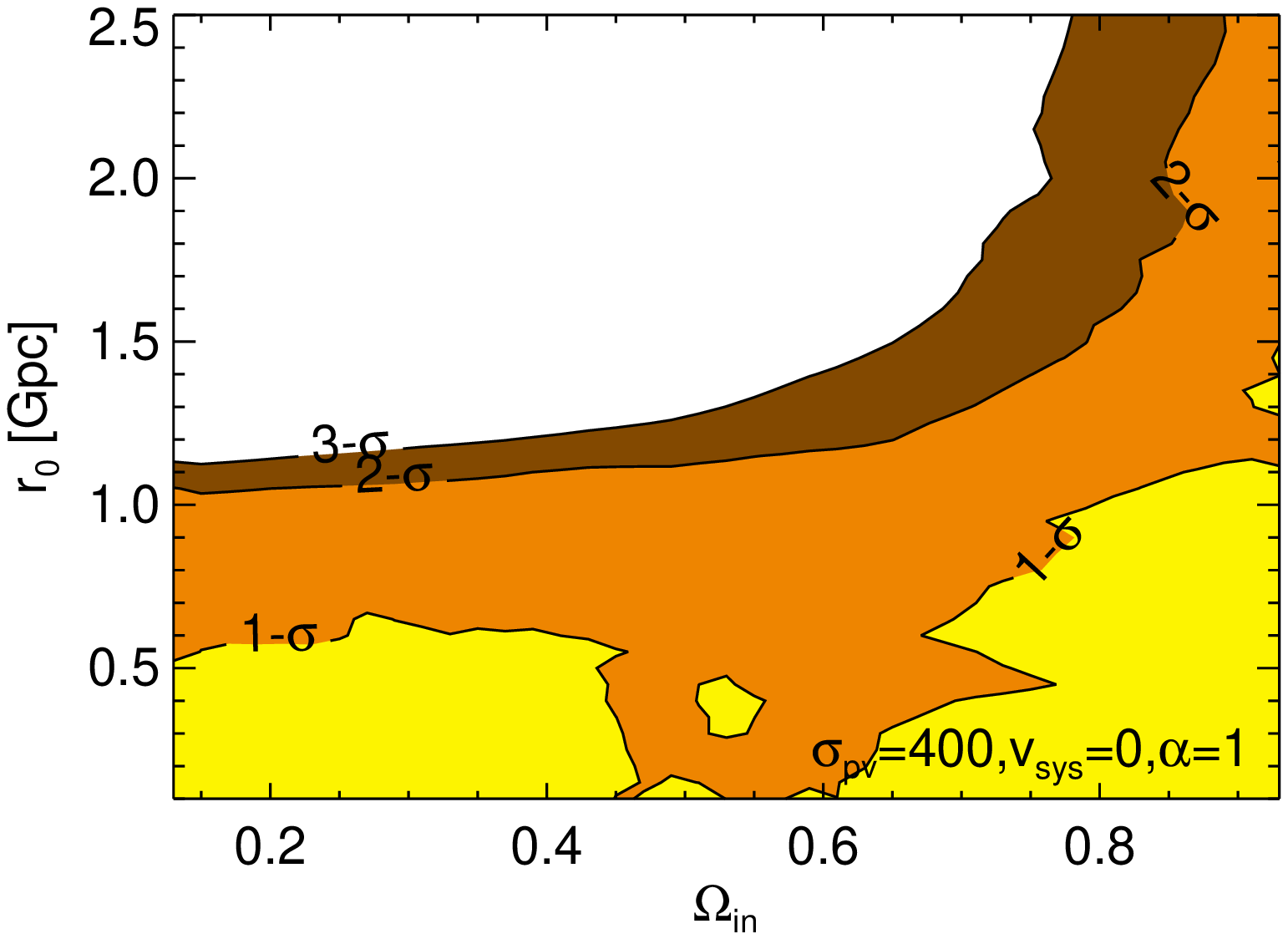}
\includegraphics[width=0.45 \textwidth]{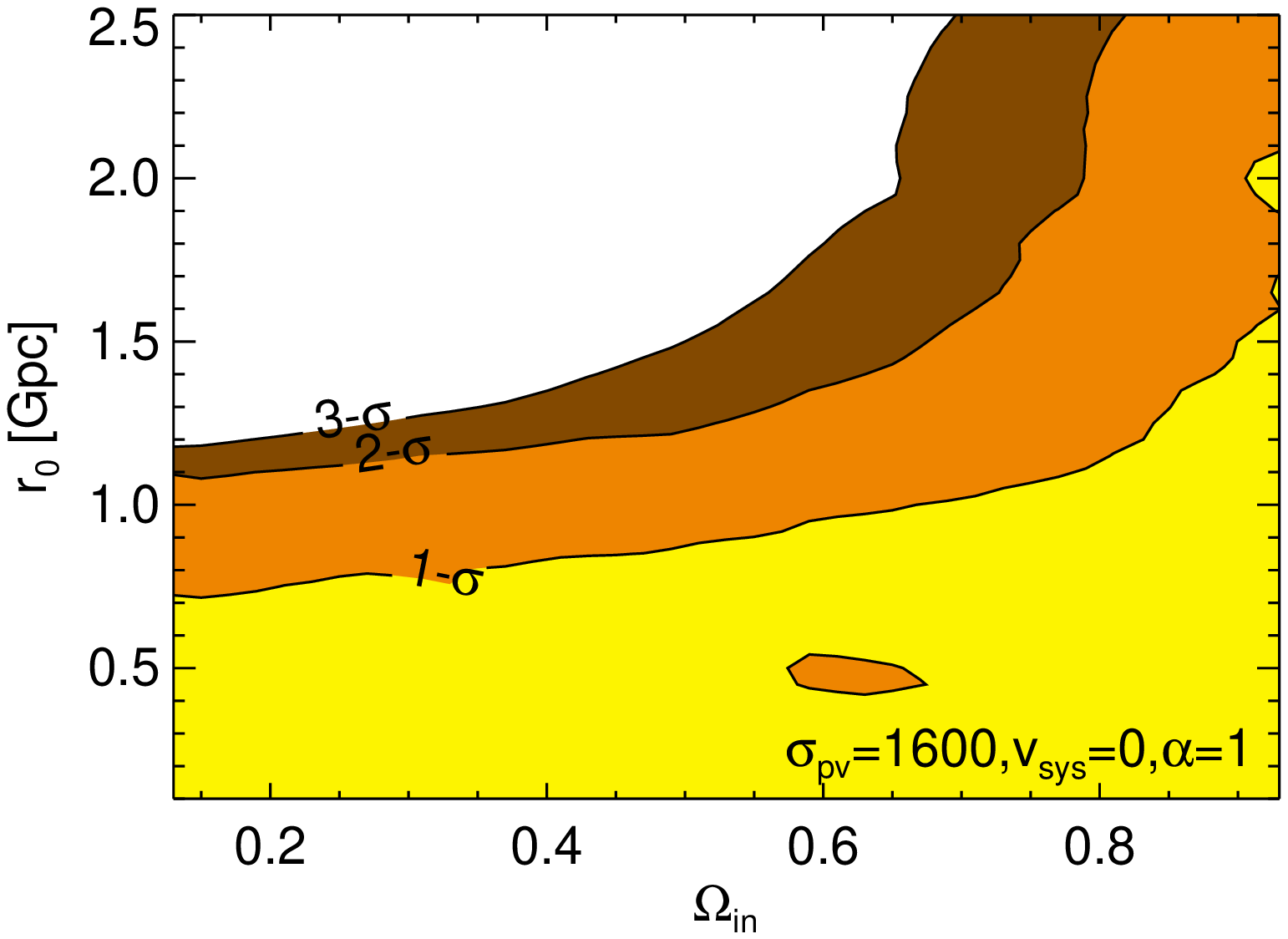} \\
\includegraphics[width=0.45 \textwidth]{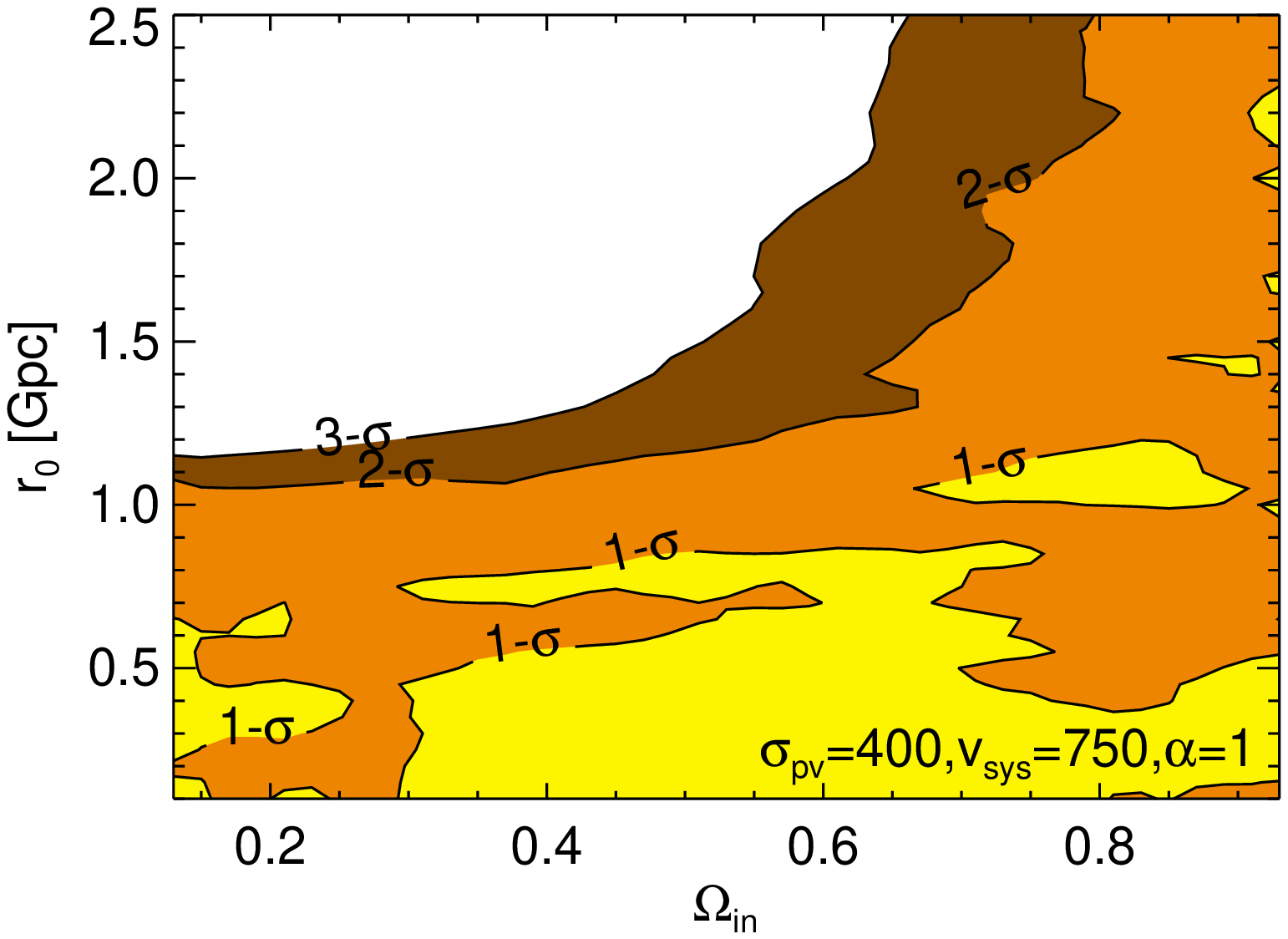}
\includegraphics[width=0.45 \textwidth]{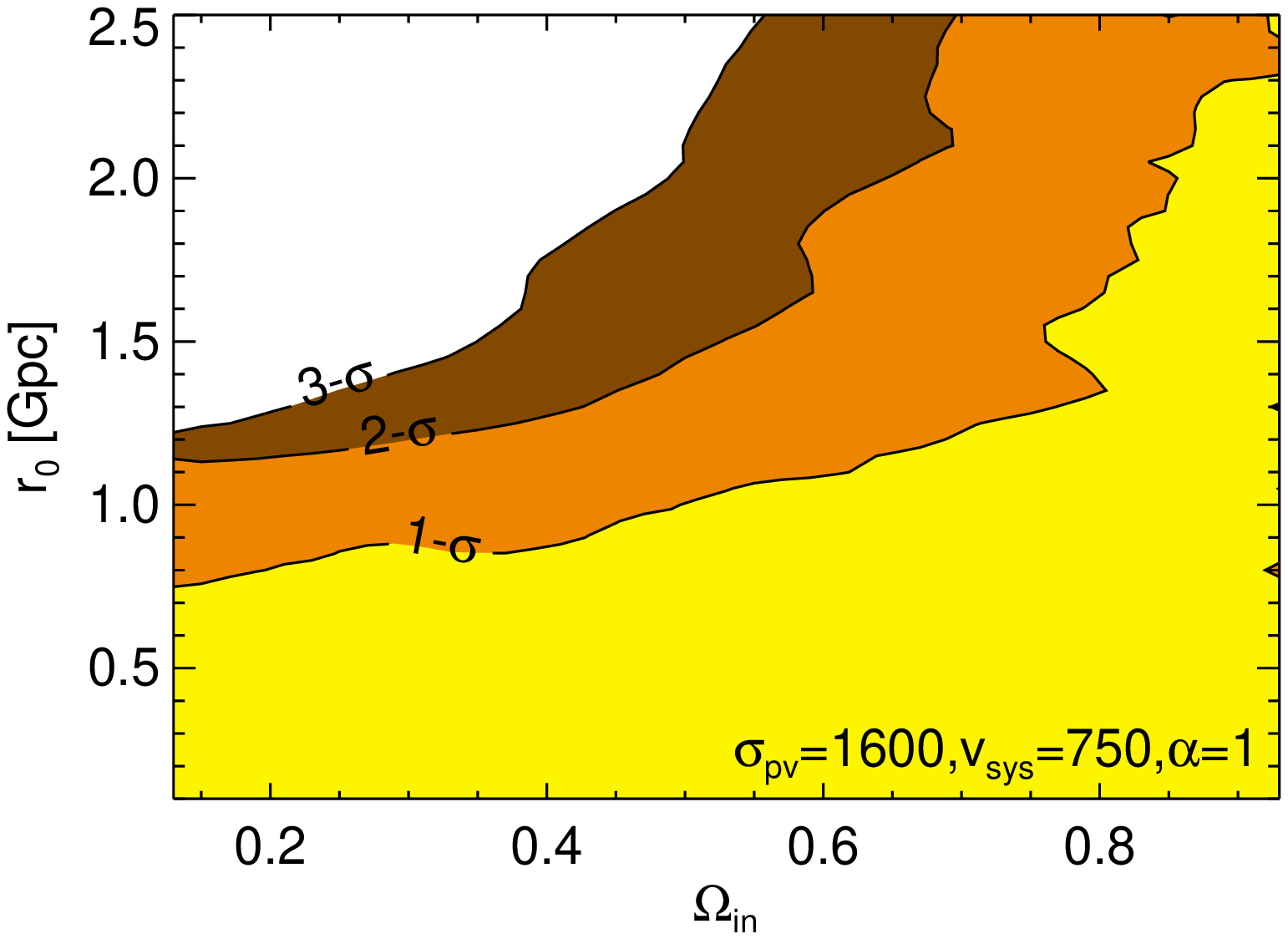} \\
\includegraphics[width=0.45 \textwidth]{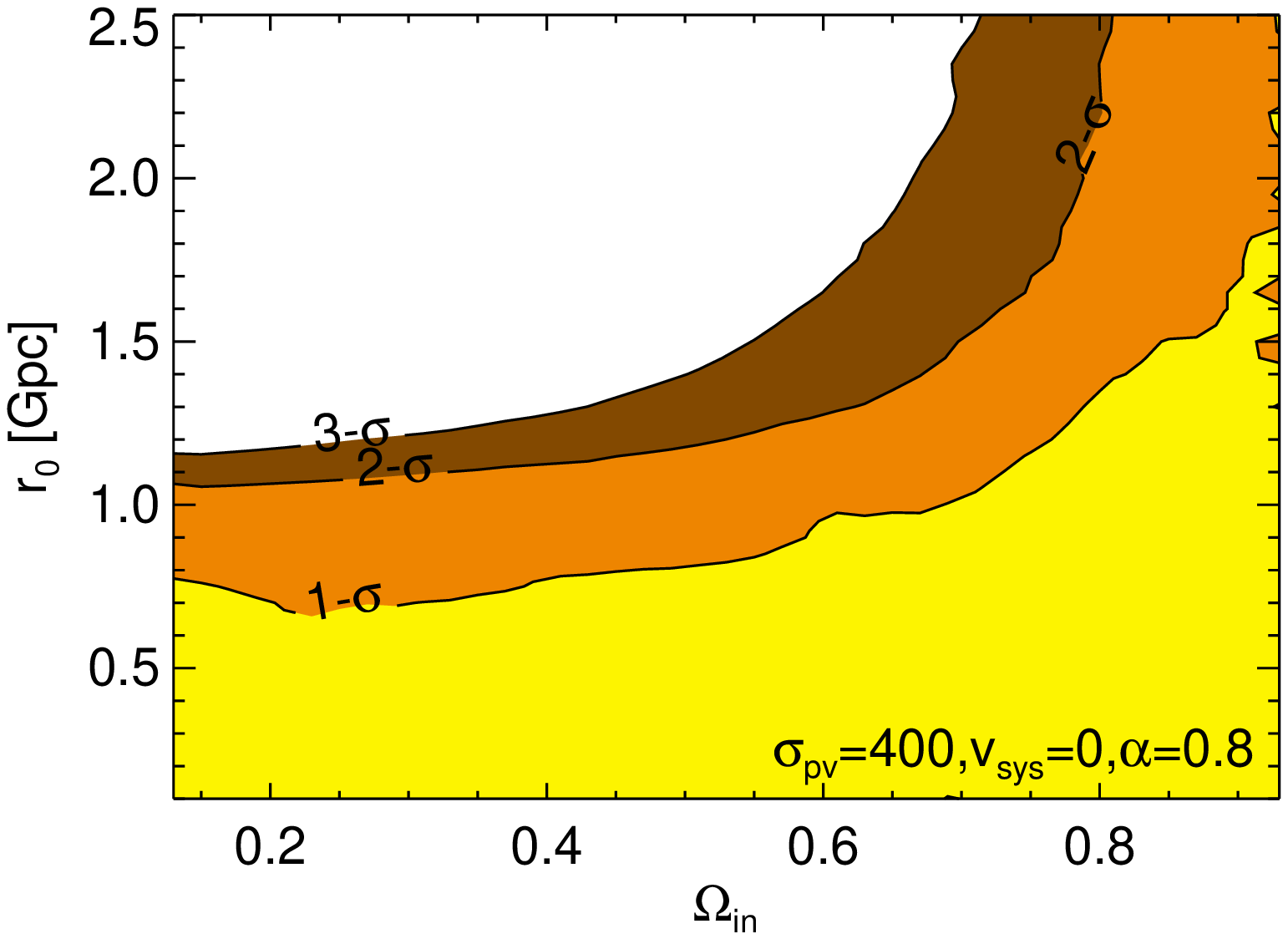}
\includegraphics[width=0.45 \textwidth]{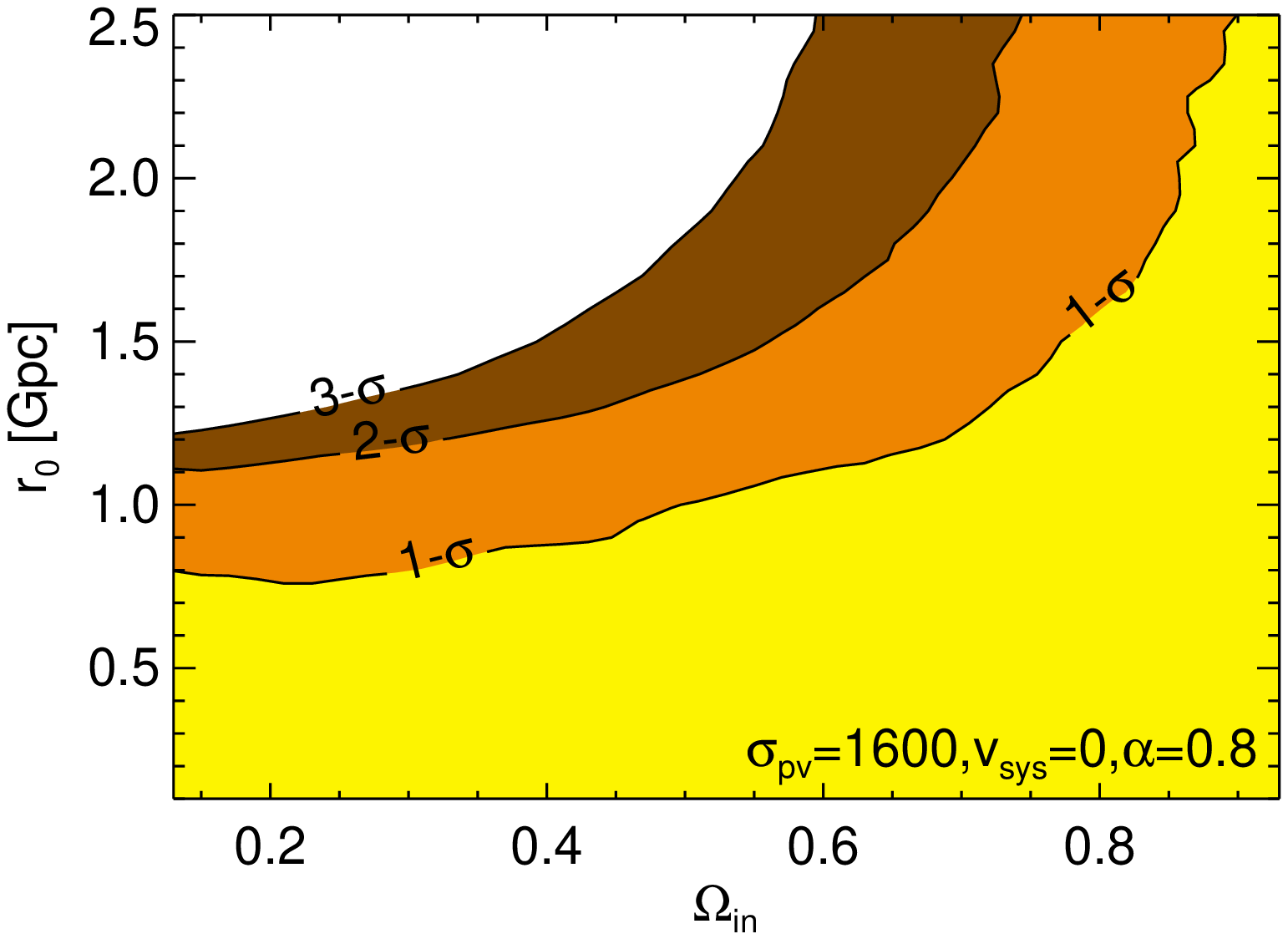} \\
\includegraphics[width=0.45 \textwidth]{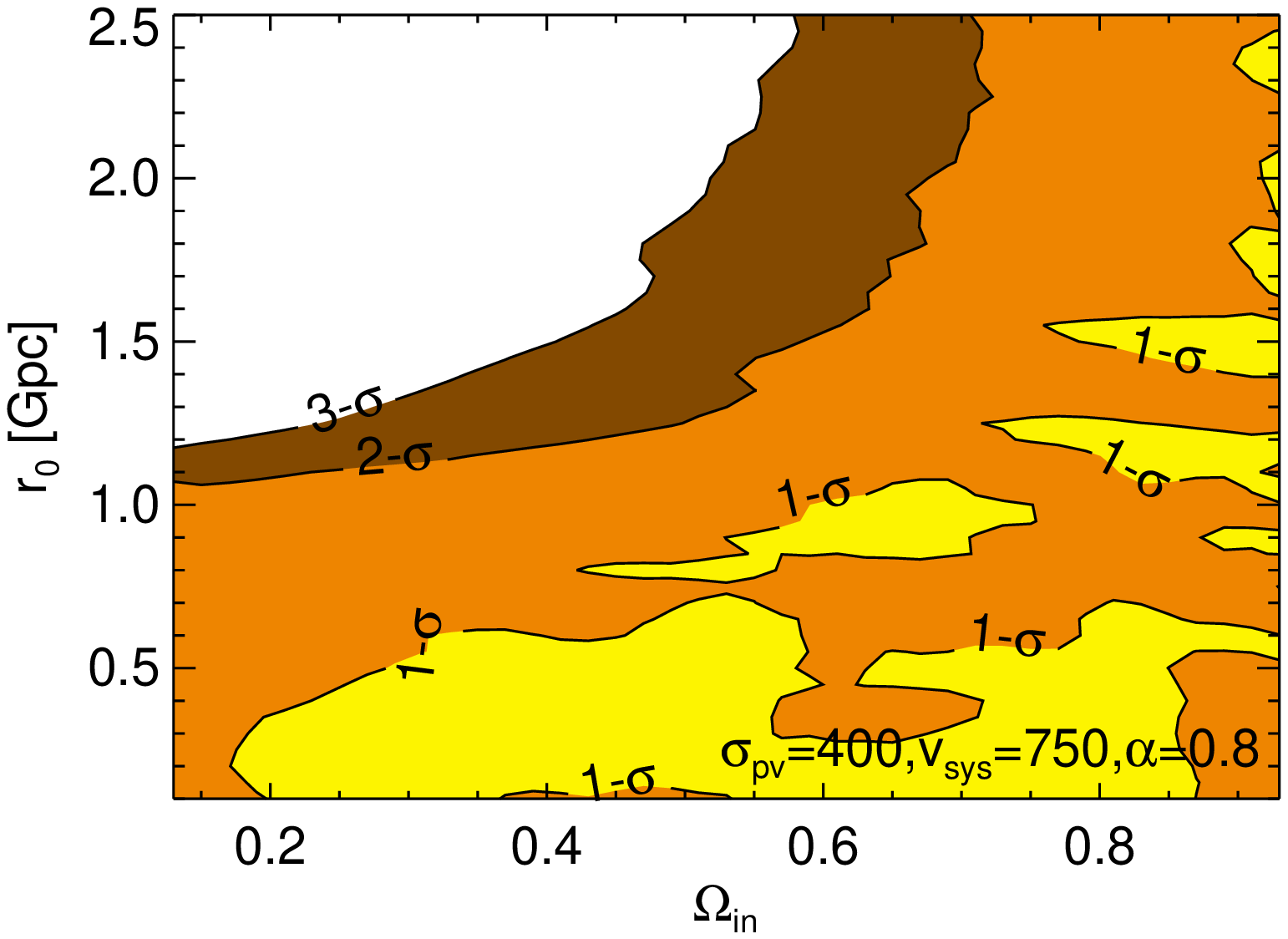}
\includegraphics[width=0.45 \textwidth]{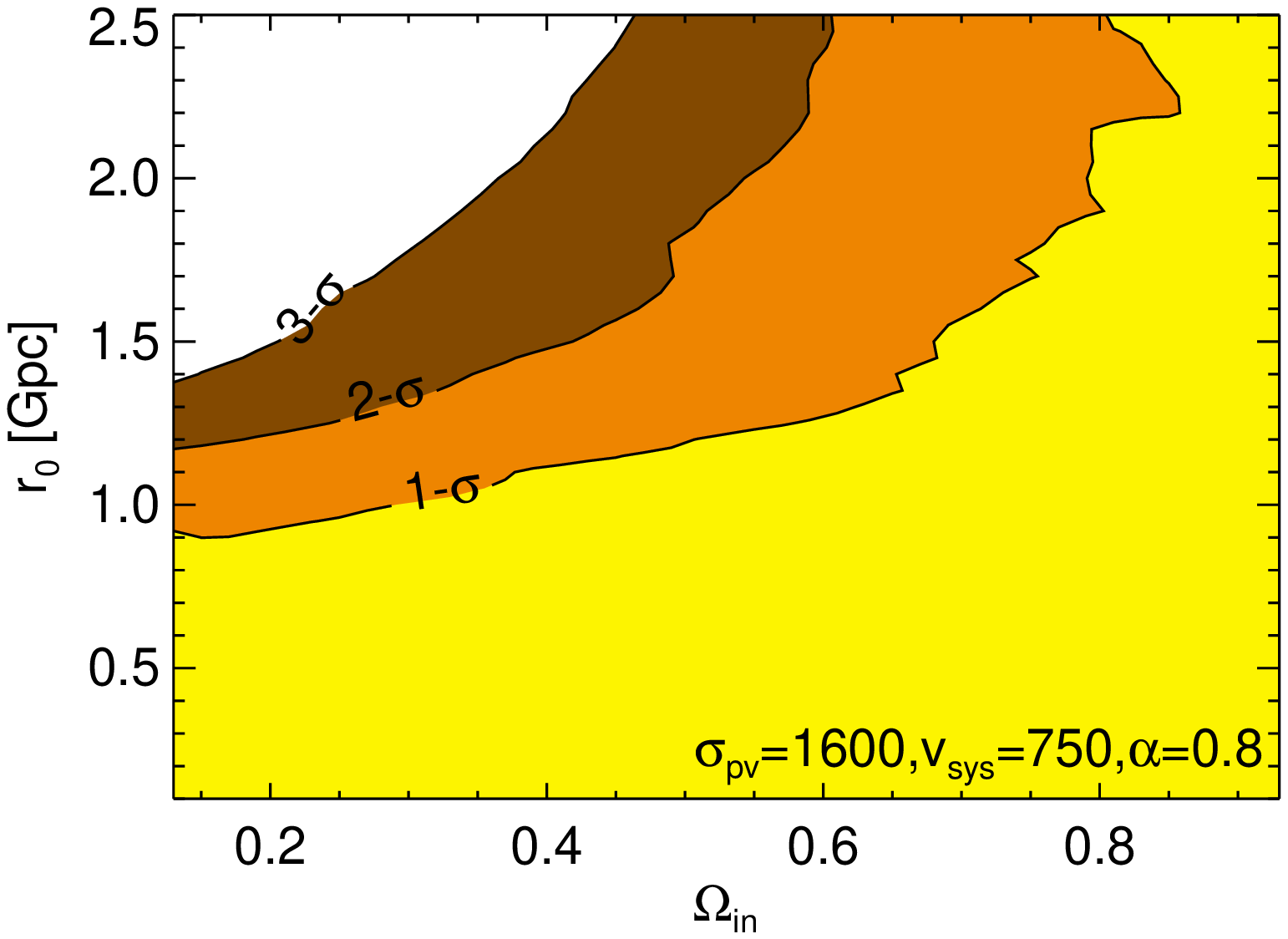}
\caption{Likelihood contours for current observed clusters with
1-$\sigma$ to 3-$\sigma$ regions given by bright to dark yellow.
From left to right, top to bottom we have $\sigma_{\rm pv} = (400, 1600)$
km s$^{-1}$, $v_{\rm sys}=(0, -750)$ km s$^{-1}$, and $\alpha=(1, 0.8)$.}\label{fig:lobs}
\end{center}
\end{figure}

\begin{figure}
\begin{center}
\includegraphics[width=0.45 \textwidth]{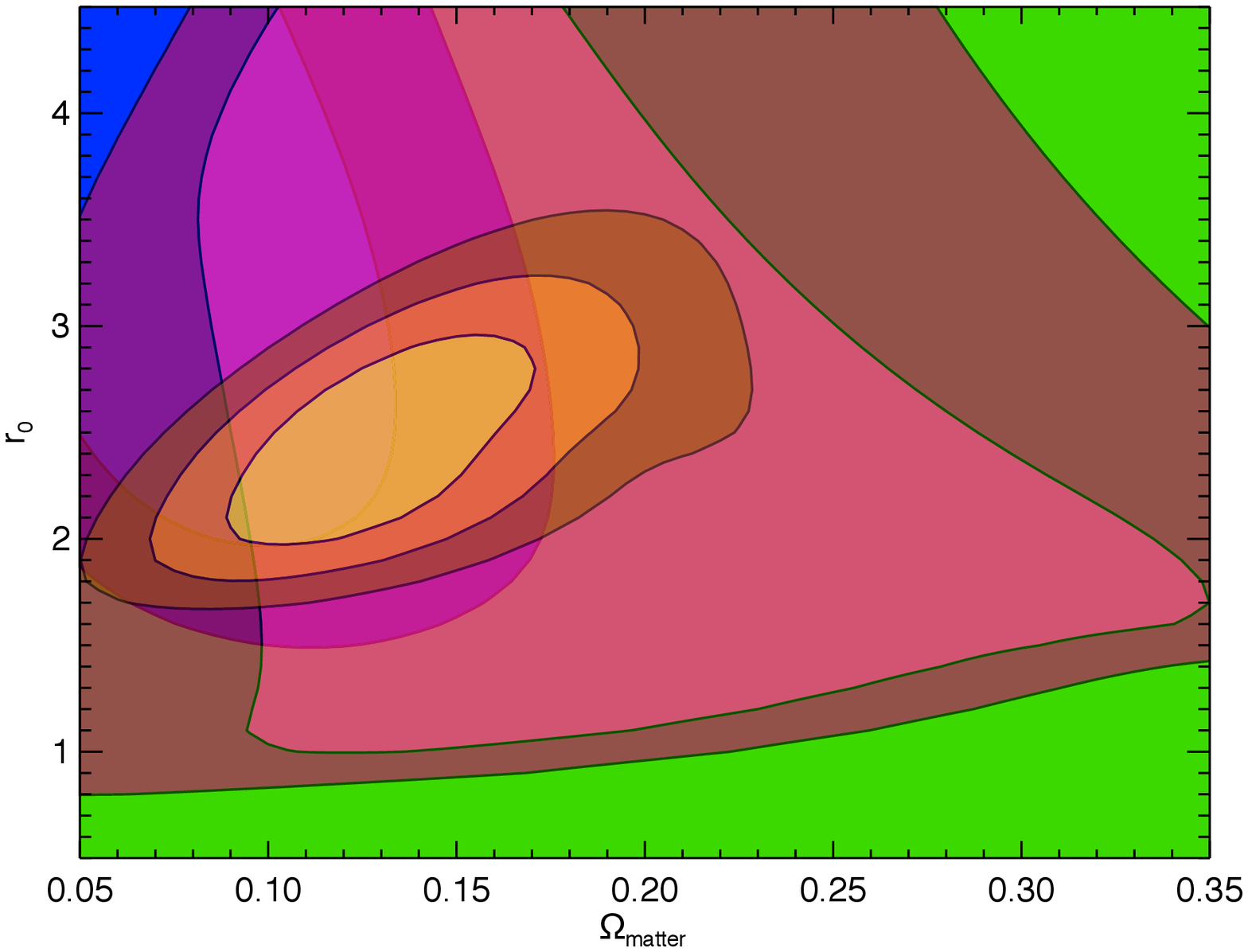}
\caption{Likelihoods for the constrained model for complementary data.
The combined data sets of Supernovae Type Ia (SNIa), BAO and WMAP
are shown in yellow with 1-, 2-, and 3-$\sigma$ contours, while the
individual SNIa, BAO, and CMB data sets are shown in blue, purple,
and green respectively with 1- and 2-$\sigma$ contours. The figure is
taken from \cite{GBH:2008}.}\label{fig:otherdata}
\end{center}
\end{figure}

The two choices of each of three free parameters leave us with 8
different likelihoods to explore. The two dimensional contours are
shown in Fig.~\ref{fig:lobs}. As can be seen in the lower right panel,
a void with a central underdensity of $\Omega_{\rm in} = 0.23$ (just
inside the 3-$\sigma$ limit allowed by other observations, see
Fig.~\ref{fig:otherdata} and Ref.~\cite{GBH:2008}), cannot be bigger
than 1.5 Gpc at the 3-$\sigma$ limit. This is the best-case, allowing
for generous errors in our modelling, an unrealistic large scatter in
the peculiar velocities, and a favourable systematic shift due to an
incomplete analysis of observations.  Incidentally, the 3-$\sigma$
limit for the size of the void, allowed by other data, is 1.45
Gpc~\cite{GBH:2008}.  For the constrained LTB model, interpreting the
current kSZ cluster data in the most favourable way, we therefore
still have at least a 3-$\sigma$ inconsistency between observations of
the geometry of the universe (Supernovae, BAO, and 1$^{\rm st}$ peak
in the CMB), and observations of the kSZ in clusters of galaxies.

\begin{table}
\begin{center}
\begin{tabular}{c|c|cc|cc}
\hline \hline
Name & Redshift & Pec vel & Syst. Error & Galactic $l$ & Galactic $b$ \\
          &                & \multicolumn{2}{c|}{[km s$^{-1}]$} & \multicolumn{2}{c}{[$^\circ$]} \\
\hline
A1689 \cite{Hopzafel:1997} & 0.18 & $+170^{+815}_{-630}$ & 750 & 313.39 & 61.10 \\
A2163 \cite{Hopzafel:1997} & 0.20 & $+490^{+1370}_{-880}$ & 750 & 6.75 & 30.52 \\
A2261 \cite{Benson:2003} & 0.22 & $-1575^{+1500}_{-975}$ & 750 & 55.61 & 31.86 \\
A2390 \cite{Benson:2003} & 0.23 & $+1900^{+6225}_{-2650}$ & 750 & 73.93 & -27.83 \\
A1835 \cite{Benson:2003} & 0.25 & $-175^{+1675}_{-1275}$ & 750 & 340.38 & 60.59 \\
Zw 3146 \cite{Benson:2003} & 0.29 & $-400^{+3700}_{-1925}$ & 750 & 239.39 & 47.96 \\
RX J1347-1145 \cite{Kitayama:2004} & 0.45 & $1420^{+1170}_{-1270}$ & ? & 324.04 & 48.81 \\
Cl 0016+16 \cite{Benson:2003} & 0.55 & $-4100^{+2650}_{-1625}$ & 750 & 201.5 & -27.32 \\
MS 0451 \cite{Benson:2003} & 0.55 & $490^{+1370}_{-880}$ & 750 & 112.55 & -45.54 \\
\hline \hline
\end{tabular} 
\end{center}
\caption{Clusters with observed velocities from the literature. The given errors are
at the 1-$\sigma$ or 68\% confidence level. The systematic error is estimated in
\cite{Benson:2003}, and is mostly due to confusion with primary CMB anisotropies,
and contributions from sub-mm point sources. These error sources are not instrument
dependent, and we add the same error to the observations in \cite{Hopzafel:1997} and
\cite{Kitayama:2004} too.}\label{tab:ksz}
\end{table}

\begin{table}
\begin{center}
\begin{tabular}{c|cc|c}
\hline \hline
Sample & $\bar v$ & $\sigma_v$ & \#clusters\\
             & \multicolumn{2}{c|}{[km s$^{-1}]$} \\
\hline
All clusters & $320^{+440}_{-400}$ & $1630^{+400}_{-350}$ & 9 \\
Cluster in \cite{Hopzafel:1997,Benson:2003} & $190^{+480}_{-430}$ & $1630^{+440}_{-370}$ & 8 \\
Clusters in \cite{Benson:2003} & $-250^{+750}_{-660}$ & $2210^{+710}_{-580}$ & 6 \\
\hline \hline
\end{tabular} 
\end{center}
\caption{Statistical properties of sub-samples, and the full sample of clusters.
The quoted values are in good agreement with what is found in
\cite{Benson:2003}.}\label{tab:sample}
\end{table}

\begin{table}[!ht]
\begin{center}
\begin{tabular}{ccc|c|cc}
\hline \hline
$H_0$ & $H_{\rm in}$ & $H_{\rm out}$ & $\Omega_{\rm in}$
           & $r_0$ & $\Delta r$ \cr
\multicolumn{3}{c|}{100 km s$^{-1}$ Mpc$^{-1}$} &
           & Gpc & $r_0$ \cr
\hline
$0.65$ & $0.44-0.57$ & $0.43$ & $0.13-0.93$
                    & $0.1-2.5$ & $0.1-0.9$ \cr
\hline \hline
\end{tabular} 
\end{center}
\caption{Priors used when scanning the parameters of the GBH models.
$H_0$ is a pre factor for $H_0(r)$ and  $H_{\rm in}$ and
$H_{\rm out}$ are derived from the priors on $\Omega_{\rm in}$ and
$H_0$.}\label{tab:priors}
\end{table}

\begin{figure}[!ht]
\begin{center}
\includegraphics[width=1.0 \textwidth]{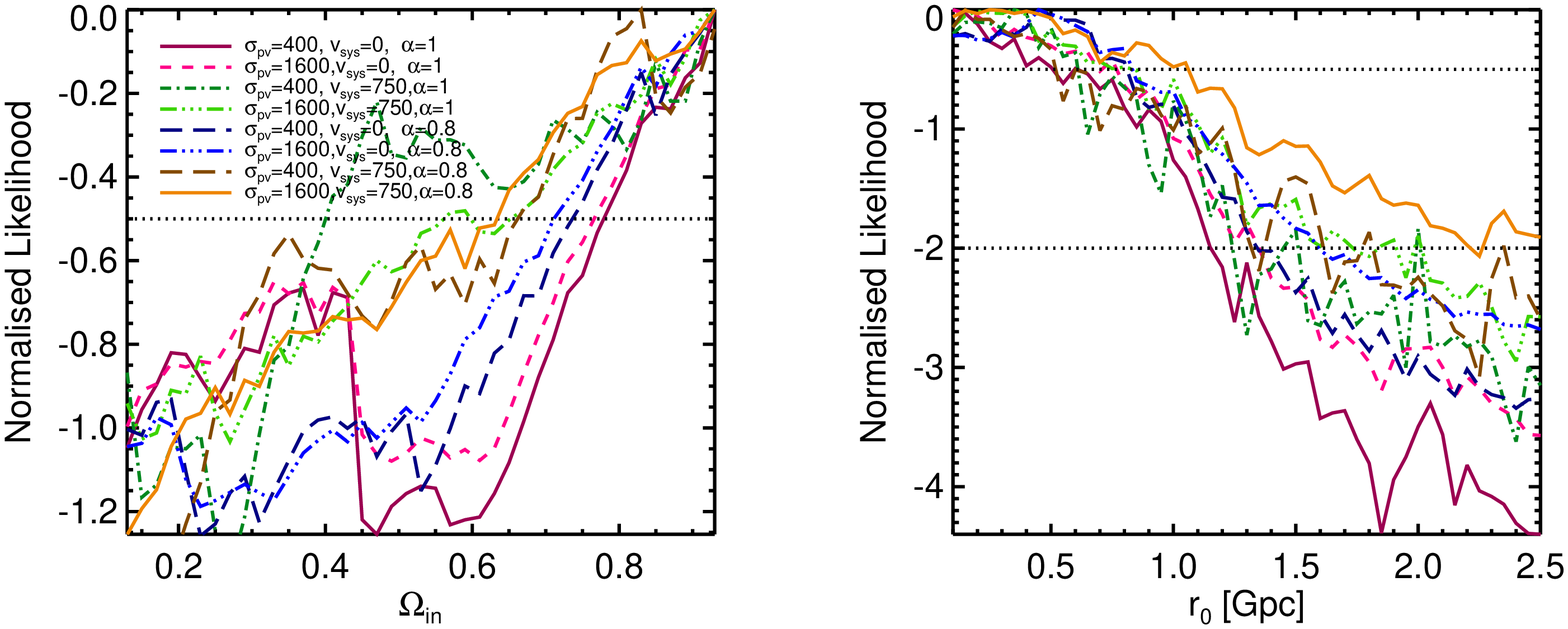}
\caption{One dimensional likelihoods for current observed clusters with
1-$\sigma$ and 2-$\sigma$ levels indicated by dotted lines. Velocities
are in km/s.}\label{fig:lobs1d}
\end{center}
\end{figure}

On the other hand, if we use the strictest interpretation of the model
(upper left panel in Fig.~\ref{fig:lobs}), then even a void with a
slight central underdensity of $\Omega_{\rm in}=0.7$ is limited to
1.5 Gpc at the 3-$\sigma$ level, or equivalently a void with a central
underdensity of $\Omega_{\rm in}<0.5$ have to be less than 1.25 Gpc
in size.

For completeness, in Fig.~\ref{fig:lobs1d} we also show the equivalent
one-dimensional likelihoods. However, we caution against
overinterpreting them: current data gives a non-detection of a local
void with four clusters moving towards us and five moving away, and
the two-dimensional likelihood contours in the $r_0$-$\Omega_{\rm in}$
plane are strongly degenerate. Hence, the reduced one-dimensional
contours give unreasonably large confidence limits, and we refrain
from quoting any general bounds on single parameters, but
conclude that measurements of the kSZ effect in clusters is by far the
most constraining way to limit the size of a local void described by
an LTB metric.

\section{Future Experiments}
In this section we explore the limits on the LTB models placed by
future experiments like the South Pole Telescope \cite{SPT}, the
Atacama Cosmology Telescope \cite{ACT}, the APEX telescope
\cite{APEX}, or the Planck satellite \cite{planck}.  Below we use
parameters for the ACT, but our results apply equally well to any of
the four surveys. Our basic hypothesis in this section is that the
Universe is homogeneous, and we show that giga parsec sized voids
with a significant underdensity are strongly ruled out, even with a
few observed galaxy clusters.

The first generation of large scale kSZ survey is characterised by a
limited number of frequency bands (ACT, APEX and SPT), or by low
angular resolution (Planck), and even though the resulting cluster
catalogues will be a gold mine for studying the thermal SZ signal, it
is much harder to exploit the full potential of the kinematic SZ for a
number of reasons: the intensity dependence is similar to that of the
primordial CMB, point sources can contaminate the fluxes, giving a
need for excellent resolution, and finally what is really measured is
the relative dip (or bump) in the frequency due to the integrated
(Thomson) scattering of photons. That is, what is measured directly is
the momentum of the cluster gas with respect to the CMB, not the
peculiar velocity itself, see Eq.~\eqref{eq:kSZ}.  If only three
(ACT) or four (SPT) frequency bands are available then an estimate of
the total gas mass, and the internal velocity dispersion, are needed
in order to extract the peculiar velocity. The gas density can be
estimated using X-rays together with thermal SZ measurements, or
alternatively using weak lensing together with extensive modelling,
while many precise redshifts of cluster galaxies in each cluster are
needed to constrain the internal velocity 
dispersion~\cite{Kosowsky:2008}.  The bottom line is that even though
$10^3$-$10^4$ clusters will be detected in the first generation of
large scale SZ surveys we can only constrain the peculiar velocities
of those clusters that are well measured by other means, severely
limiting the clusters that can be explored, to those at lower
redshifts ($z \lesssim 0.4$). Hence the number of clusters with kSZ
measurements will be close to the number of clusters which are well
observed in X-rays, i.e. $10^2$-$10^3$ clusters.

\begin{figure}
\begin{center}
\includegraphics[width=0.55 \textwidth]{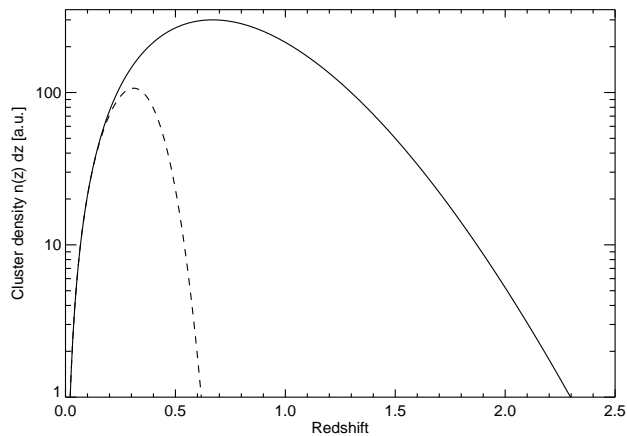}
\caption{Cluster redshift number density used to select the redshift
  distribution for forecasting future surveys. The dashed line is the
  number density folded with the X-ray cut-off.}\label{fig:nz}
\end{center}
\end{figure}

To construct a mock survey we model the cluster PDF as
\begin{equation}\label{eq:nz}
n(z) \propto \left( \frac{z}{z_{\rm max}} \right)^2
  \exp\left[-\left({z\over z_{\rm max}}\right)^{5/3}\right] 
  \exp\left[-\left({z\over z_X}\right)^4\right]\,,
\end{equation}
in accordance with the cluster density 
given in \cite{SPT} if we set $z_{\rm max}=0.6$, see Fig.~\ref{fig:nz}. 
The last term is an exponential redshift cut-off, due to the limited
reach of X-ray observations, and we set $z_X=0.4$. Notice that the
detailed form of the cluster PDF is not so important because X-ray
observations mostly sample redshifts lower than the maximum where the
volume density is nearly constant, and the redshift density then goes
like $z^2$.

We use a similar likelihood function for our mock survey as the one in
Eq.~\eqref{eq:lobs}
\begin{equation}\label{eq:lmock}
-2 \ln \mathcal{L} = \sum_{i} \left[\frac{v_i - \alpha v(\vec p,z_i) 
                     + v_{\rm sys}}{[\sigma^2_i + \sigma^2_{\rm pv}]^{1/2}}\right]^2\,,
\end{equation}
where the only difference is that we assume Gaussian errors for our
observations. Due to uncertainties in the internal velocity scatter in
the cluster gas, confusion with the primordial CMB and confusion with
point sources in any future experiment it will be very difficult to
reduce the uncertainty in the peculiar velocity $\sigma_i$ to less
than $\sim 100$ km s$^{-1}$ \cite{Kosowsky:2008}.  In the first
generation experiments it will be hard to get an uncertainty
$\sigma_i \lesssim 500$ km s$^{-1}$, due to the limited number of
frequency bands, uncertainties in measuring the gas mass by other
means, and incomplete knowledge of point sources in the clusters
\cite{Kosowsky:2008}. Below we will consider a best-case scenario of
$\sigma_i=400$ km s$^{-1}$, and a worst case of $\sigma_i=800$ km
s$^{-1}$. For simplicity we will fix the rescaling of the modelled
velocities due to uncertainties in the theoretical model to
$\alpha = 0.8$, and we will consider systematic errors of $v_{\rm
  sys}=$ 0 and 400 km s$^{-1}$, while the scatter in the peculiar
velocities is fixed as above to $\sigma_{\rm pv}=400$ and 1600 km
s$^{-1}$. These are reasonable choices that covers the probable spread
in both the observable and the theoretically motivated parameters in
Eq.~\eqref{eq:lmock}.

\begin{figure}
\begin{center}
\includegraphics[width=0.45 \textwidth]{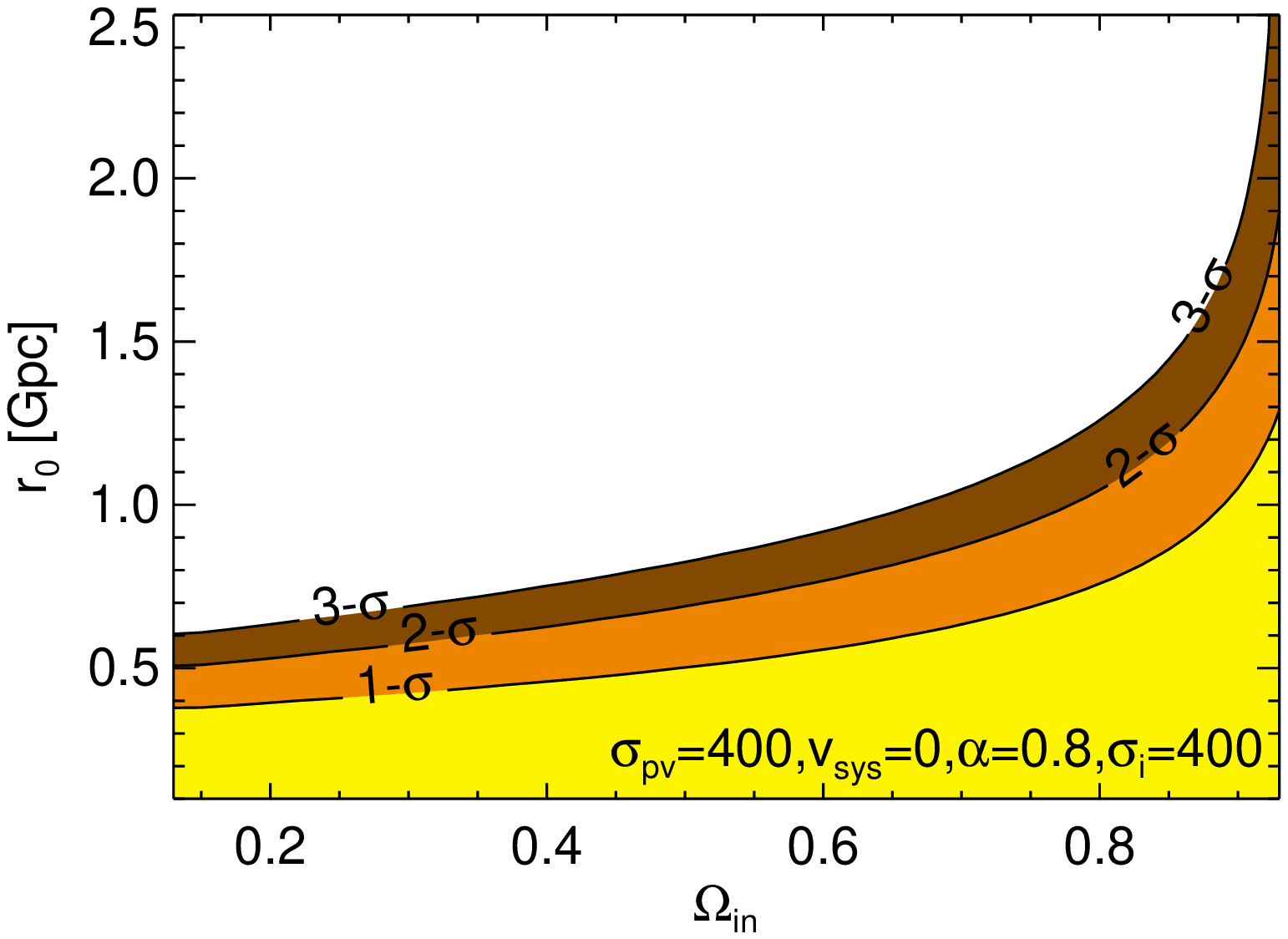}
\includegraphics[width=0.45 \textwidth]{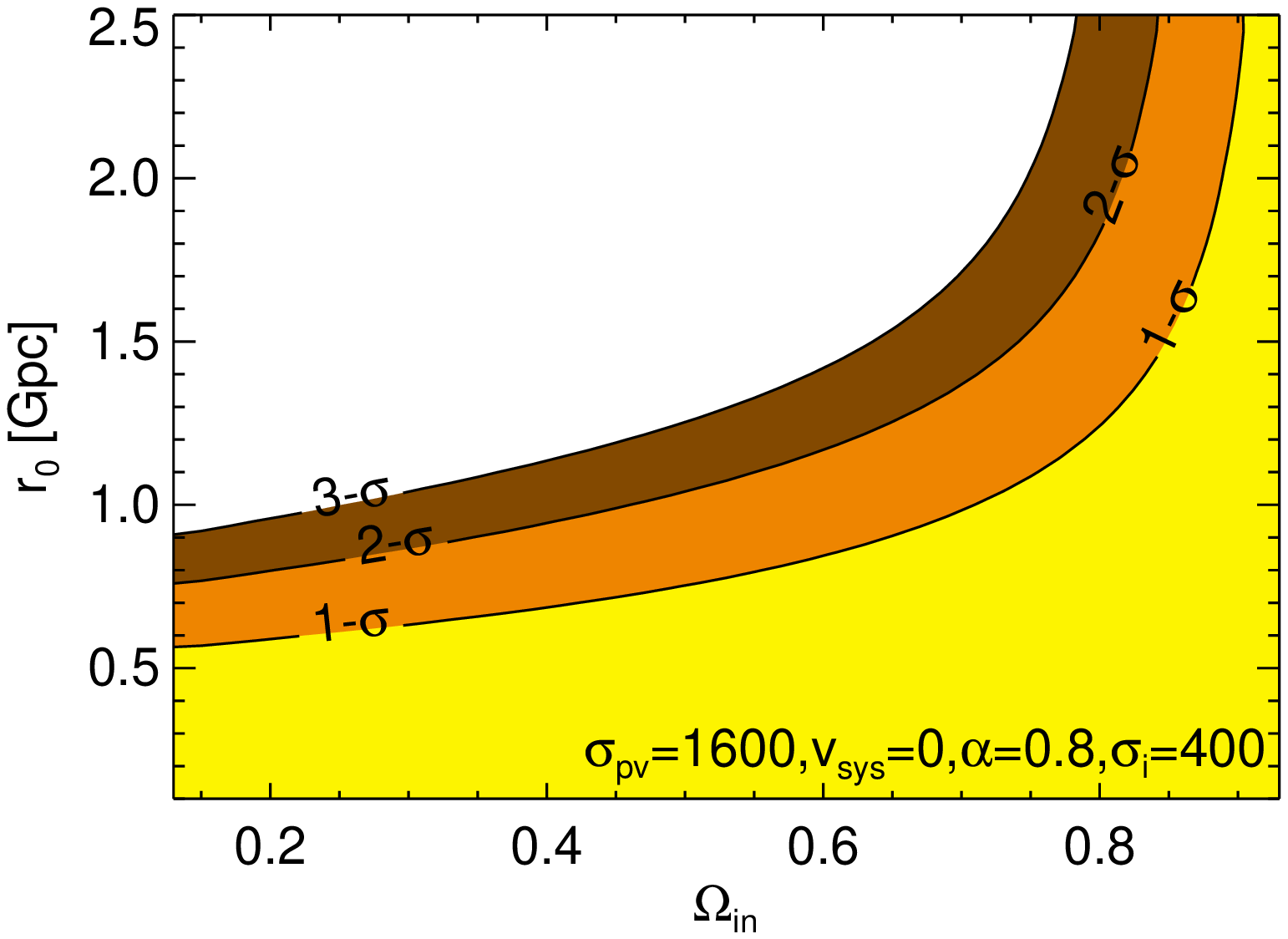} \\
\includegraphics[width=0.45 \textwidth]{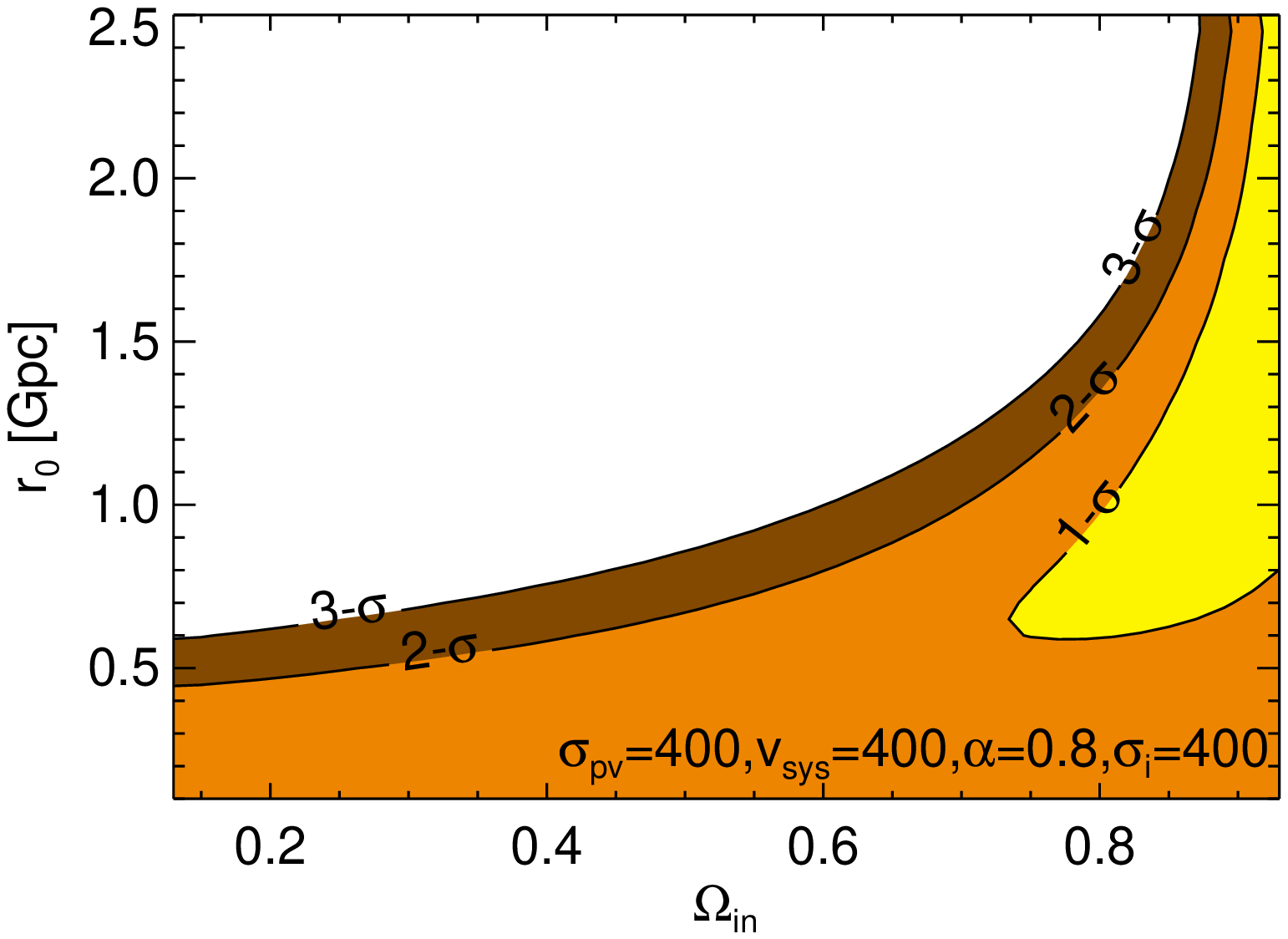}
\includegraphics[width=0.45 \textwidth]{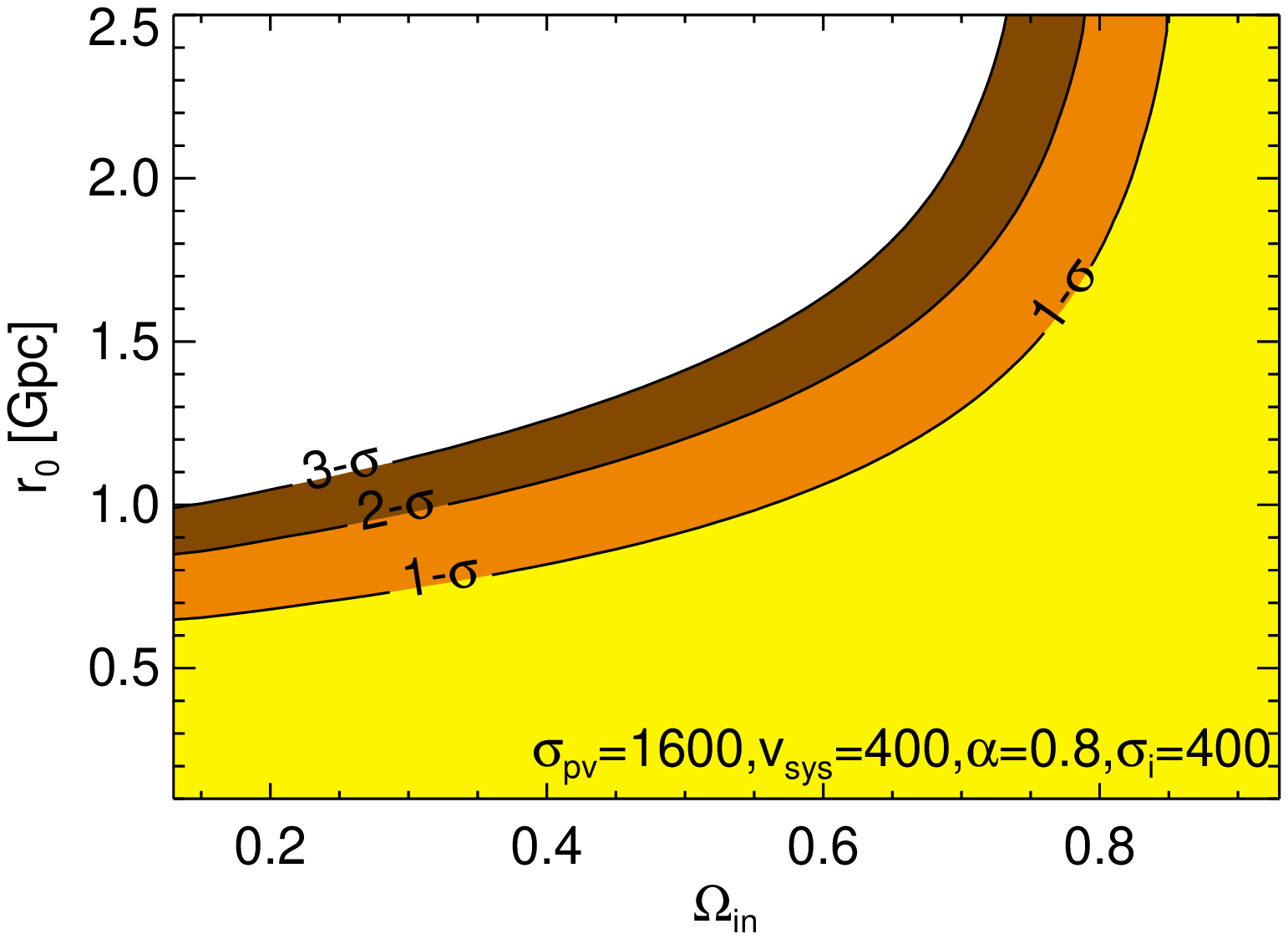} \\
\includegraphics[width=0.45 \textwidth]{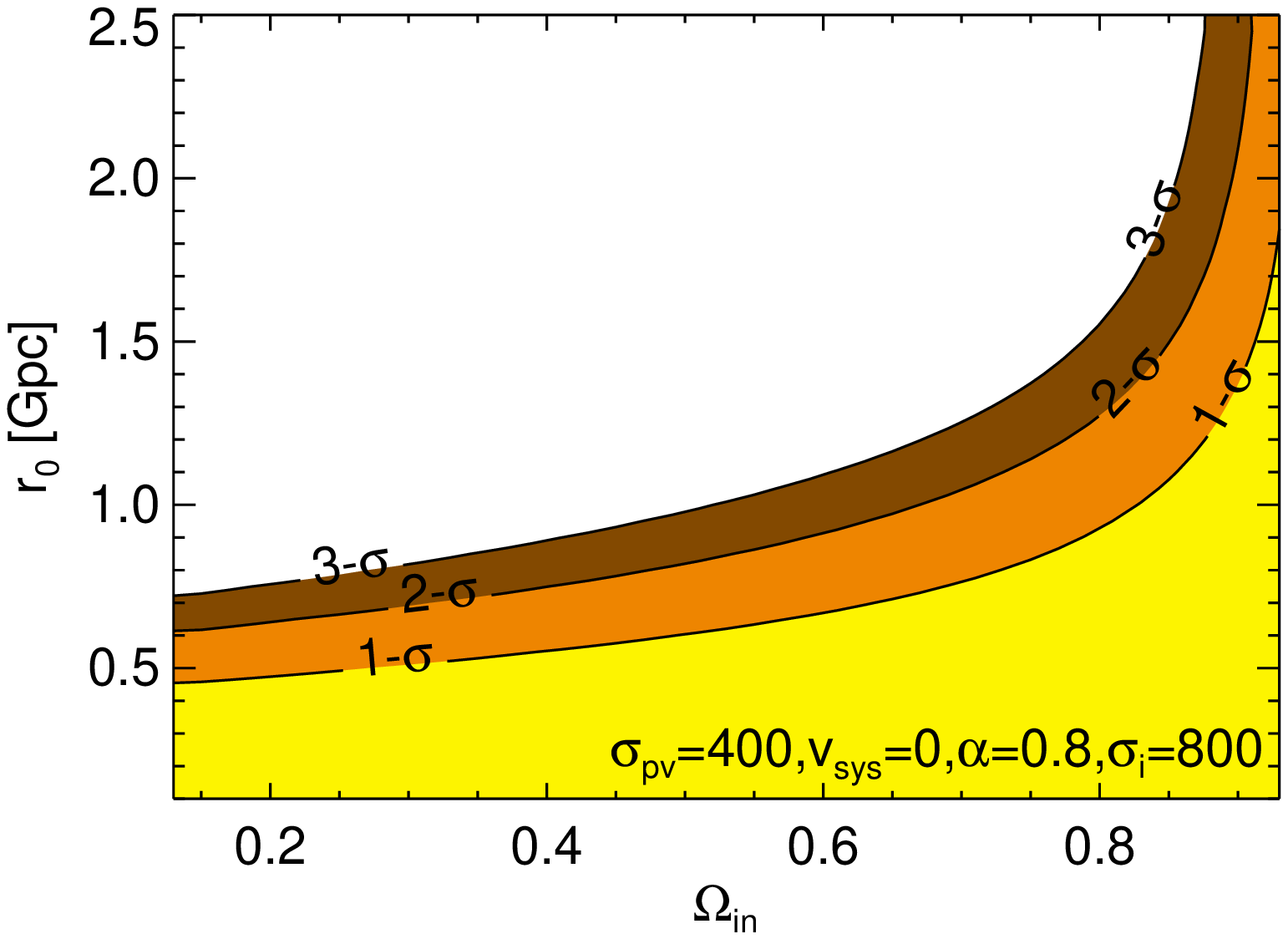}
\includegraphics[width=0.45 \textwidth]{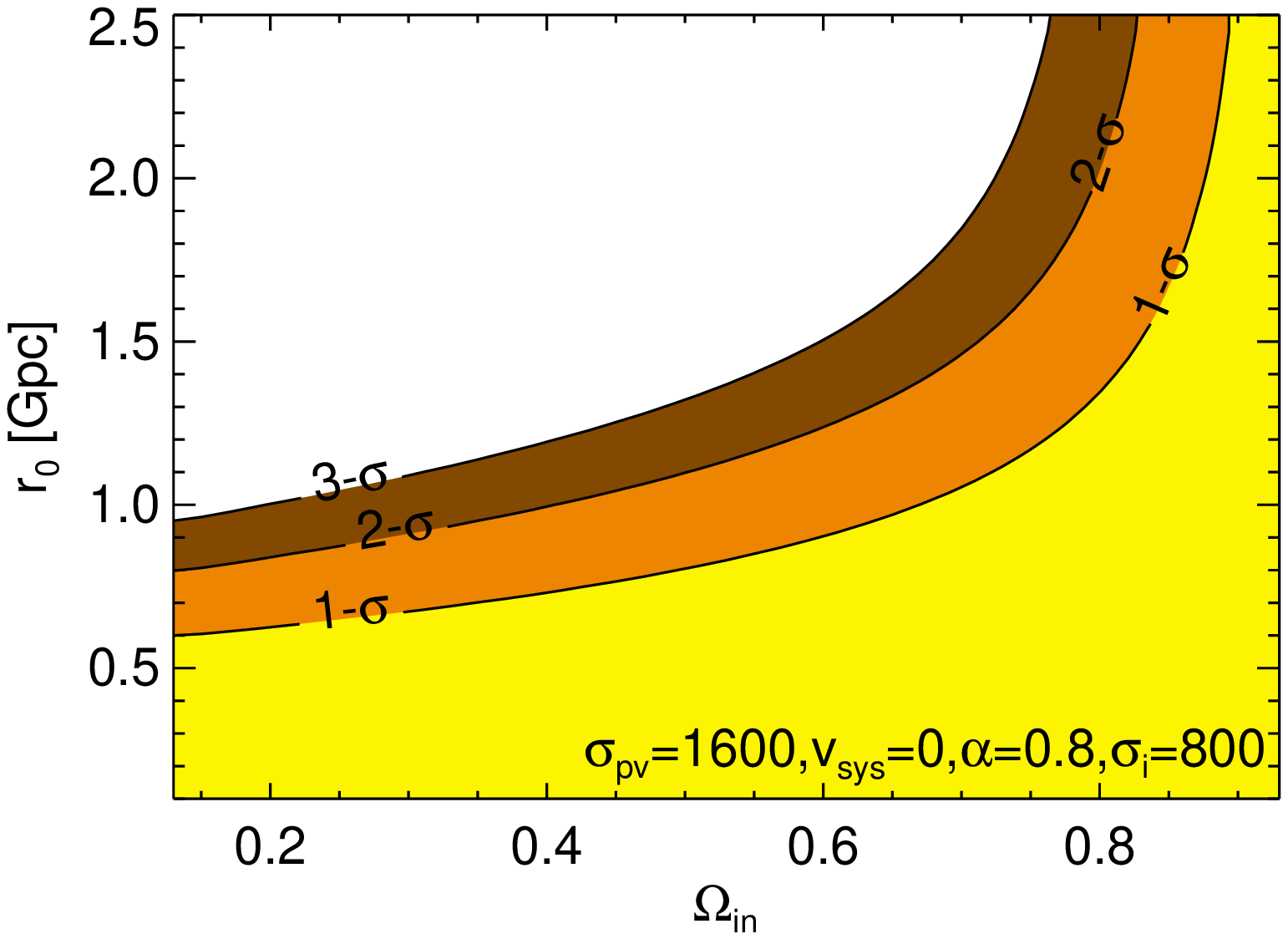} \\
\includegraphics[width=0.45 \textwidth]{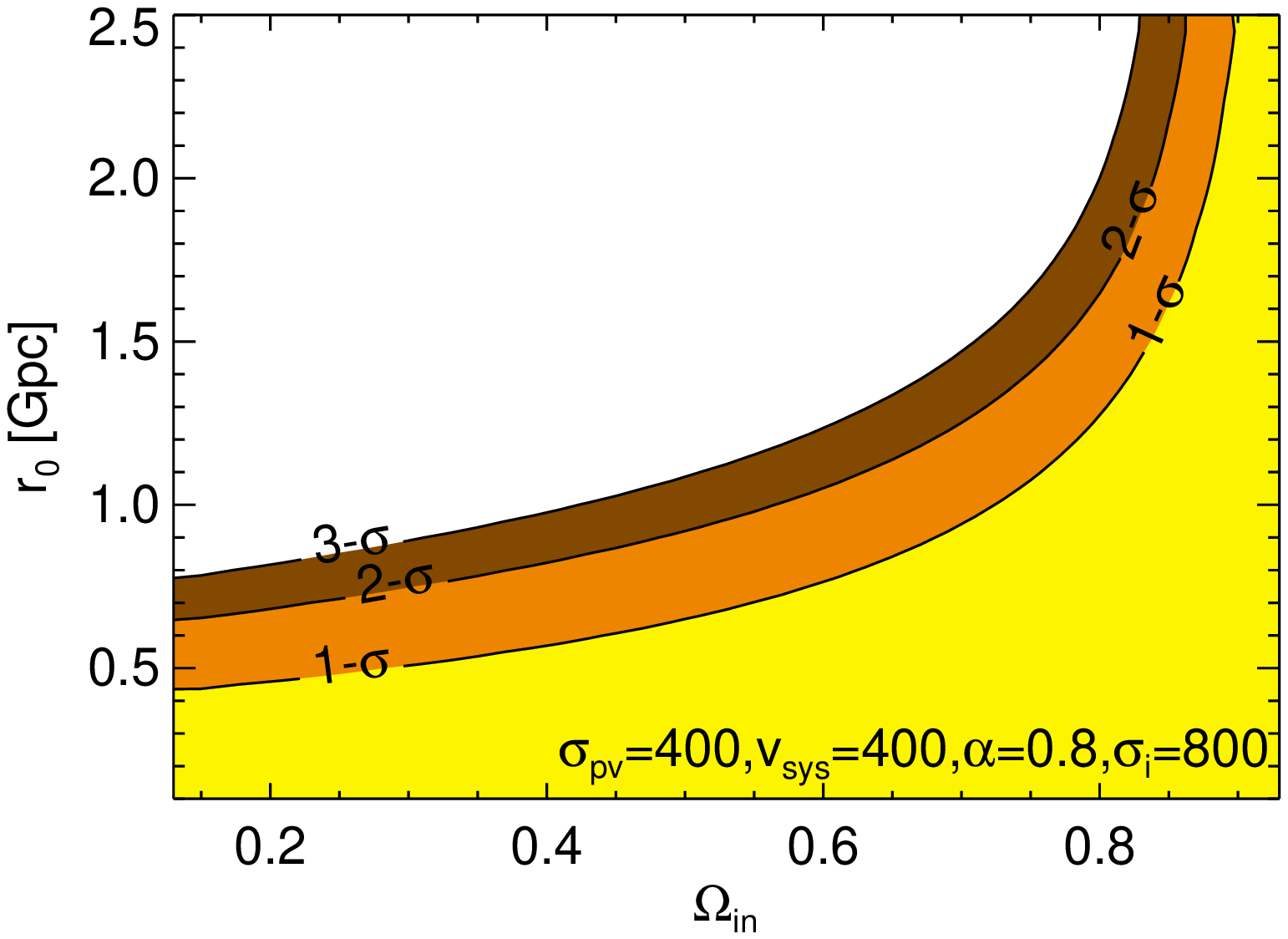}
\includegraphics[width=0.45 \textwidth]{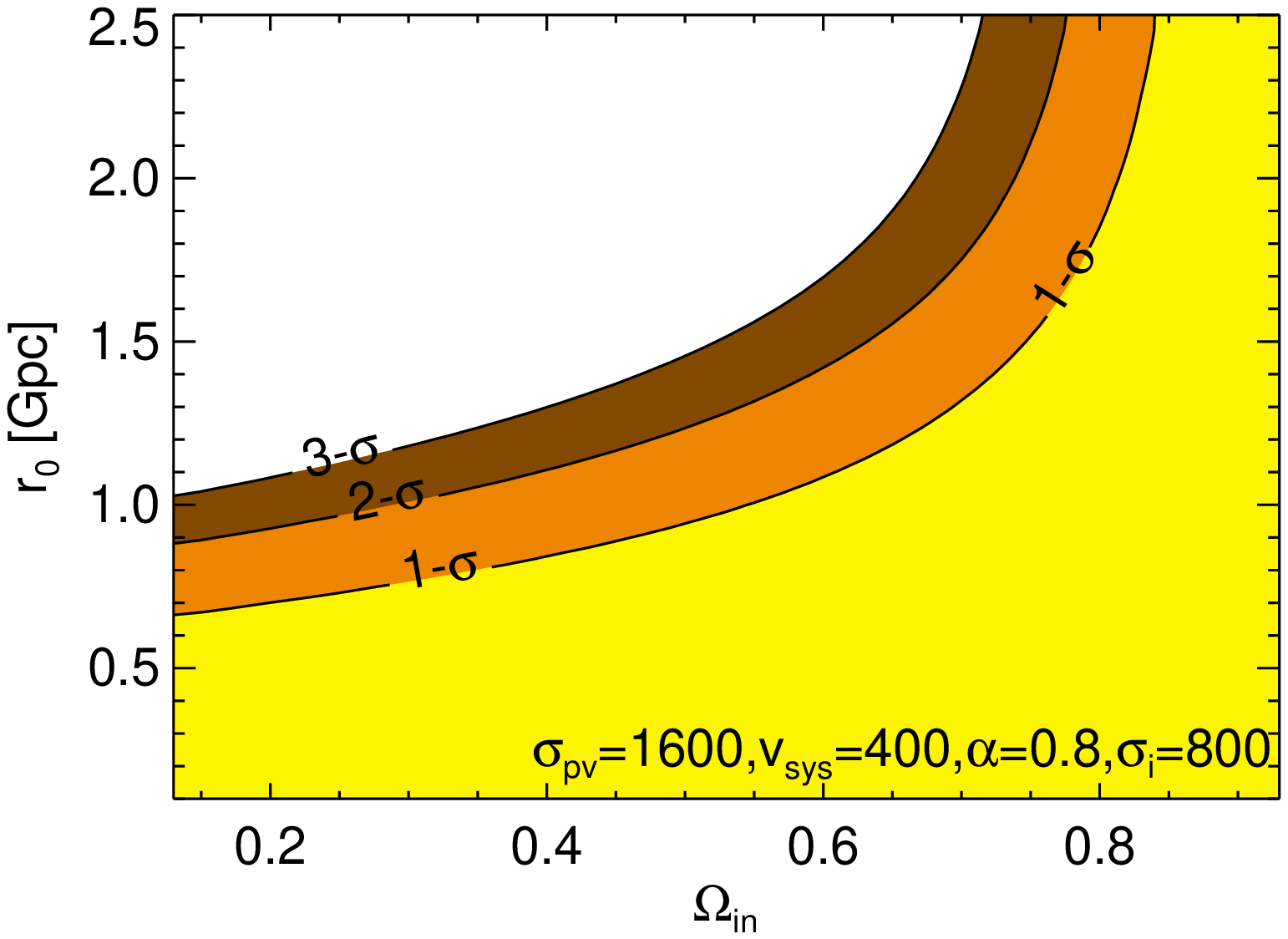}
\caption{Likelihood contours for a mock survey with 10 well observed
  clusters with 1-$\sigma$ to 3-$\sigma$ regions given by bright to
  dark yellow. Also indicated in the individual plots are the parameters
  used for the simulation. Velocities are given in km s$^{-1}$.}\label{fig:forecastnc10}
\end{center}
\end{figure}

\begin{figure}
\begin{center}
\includegraphics[width=0.45 \textwidth]{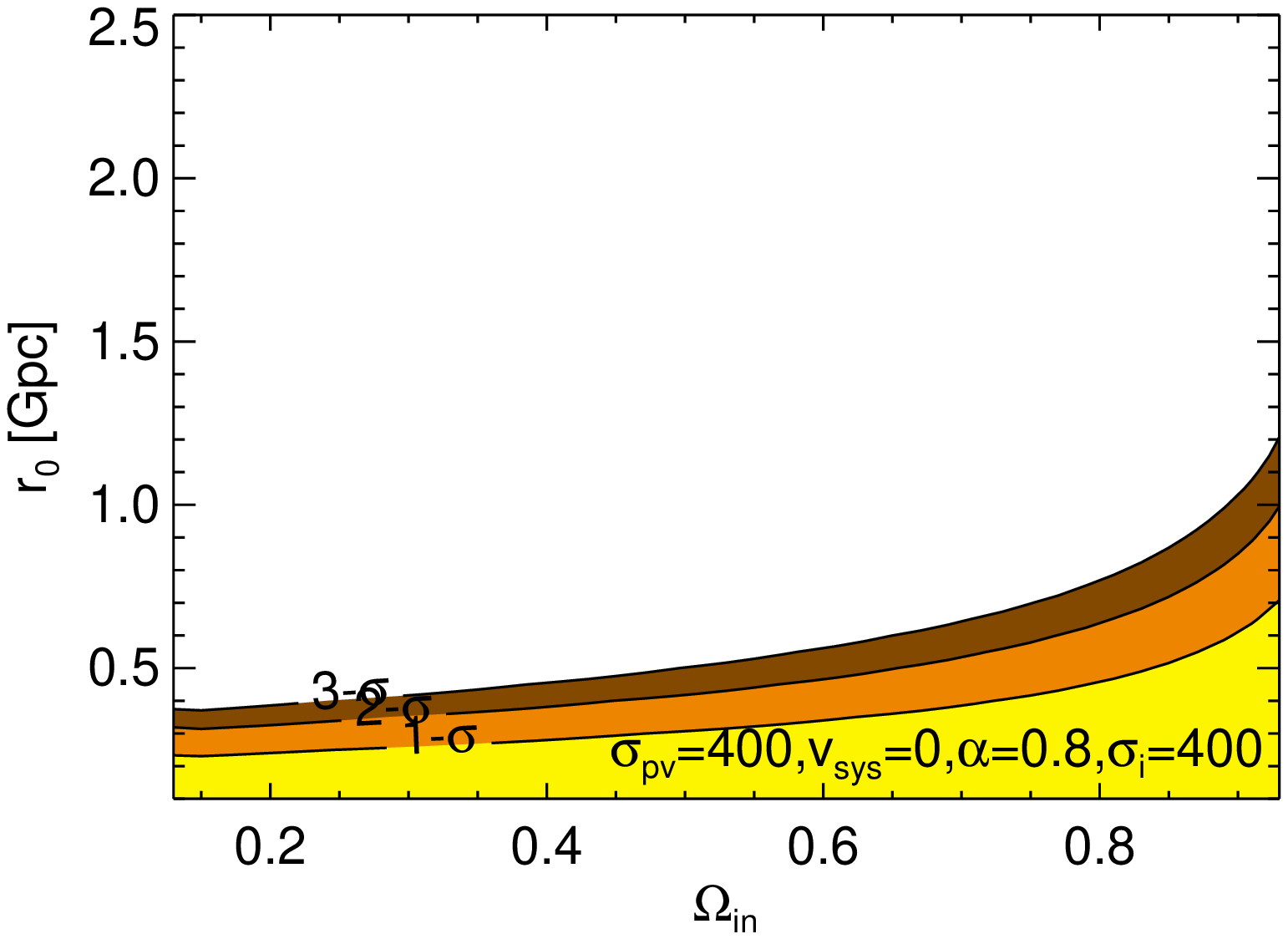}
\includegraphics[width=0.45 \textwidth]{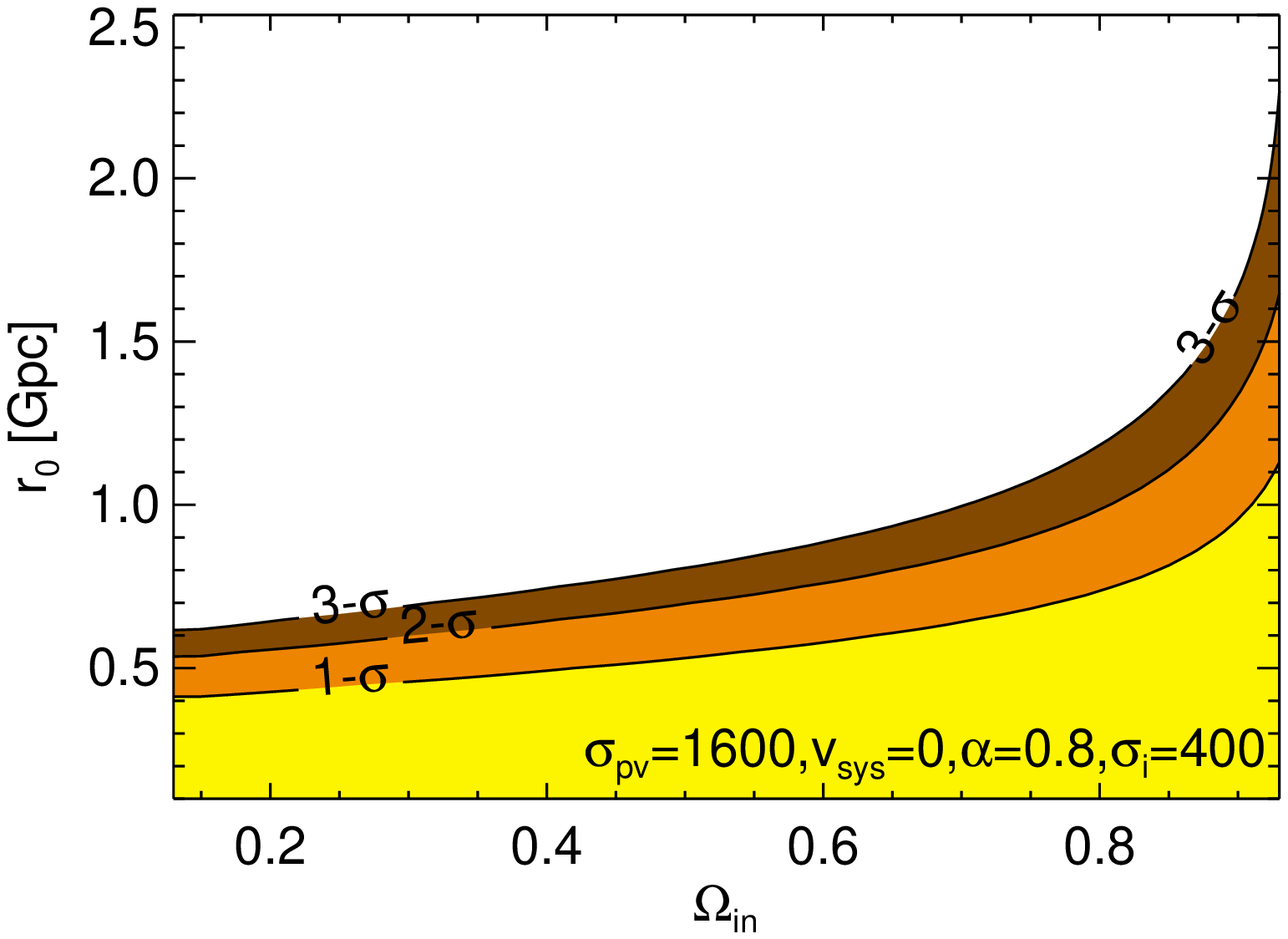} \\
\includegraphics[width=0.45 \textwidth]{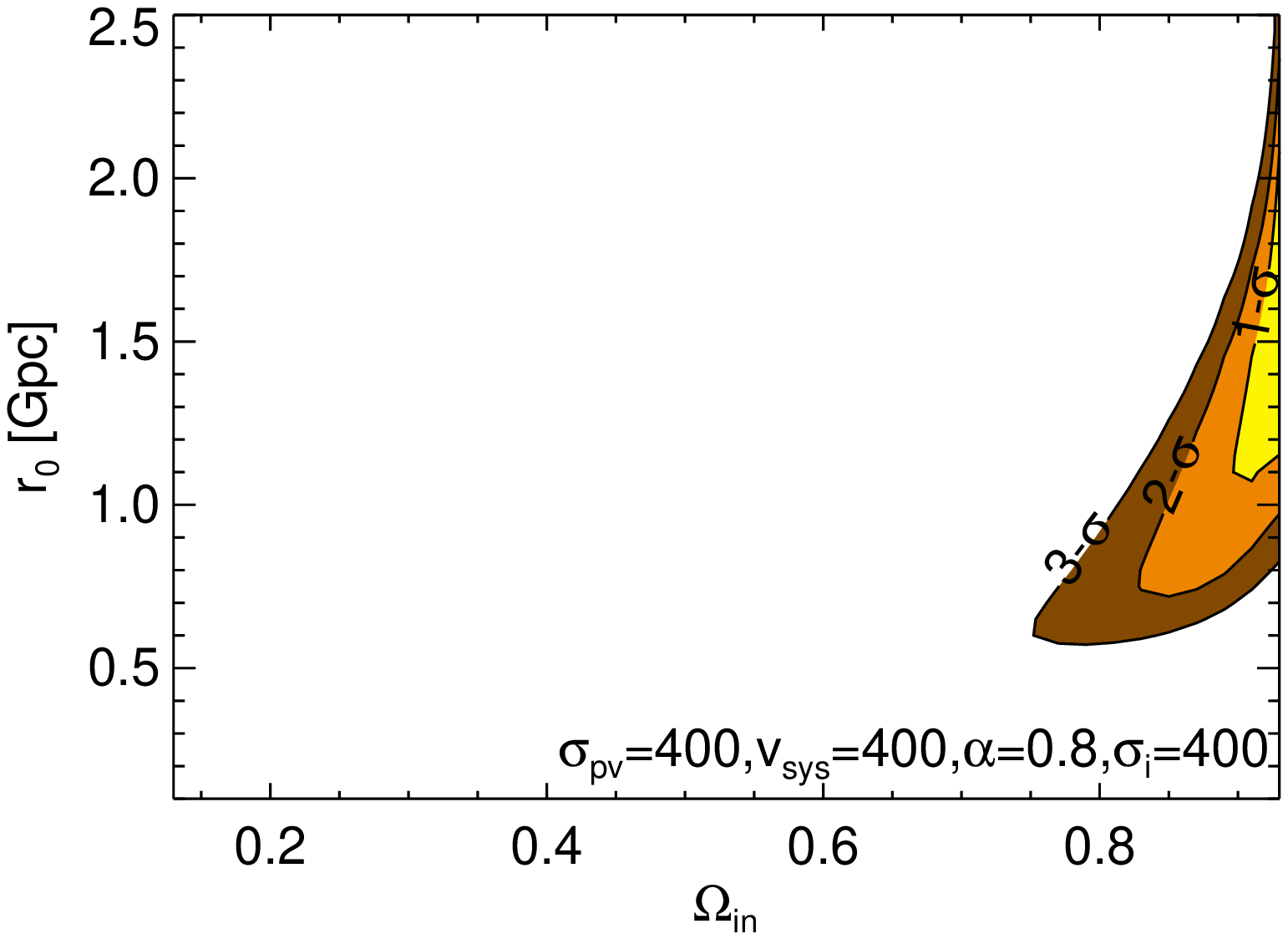}
\includegraphics[width=0.45 \textwidth]{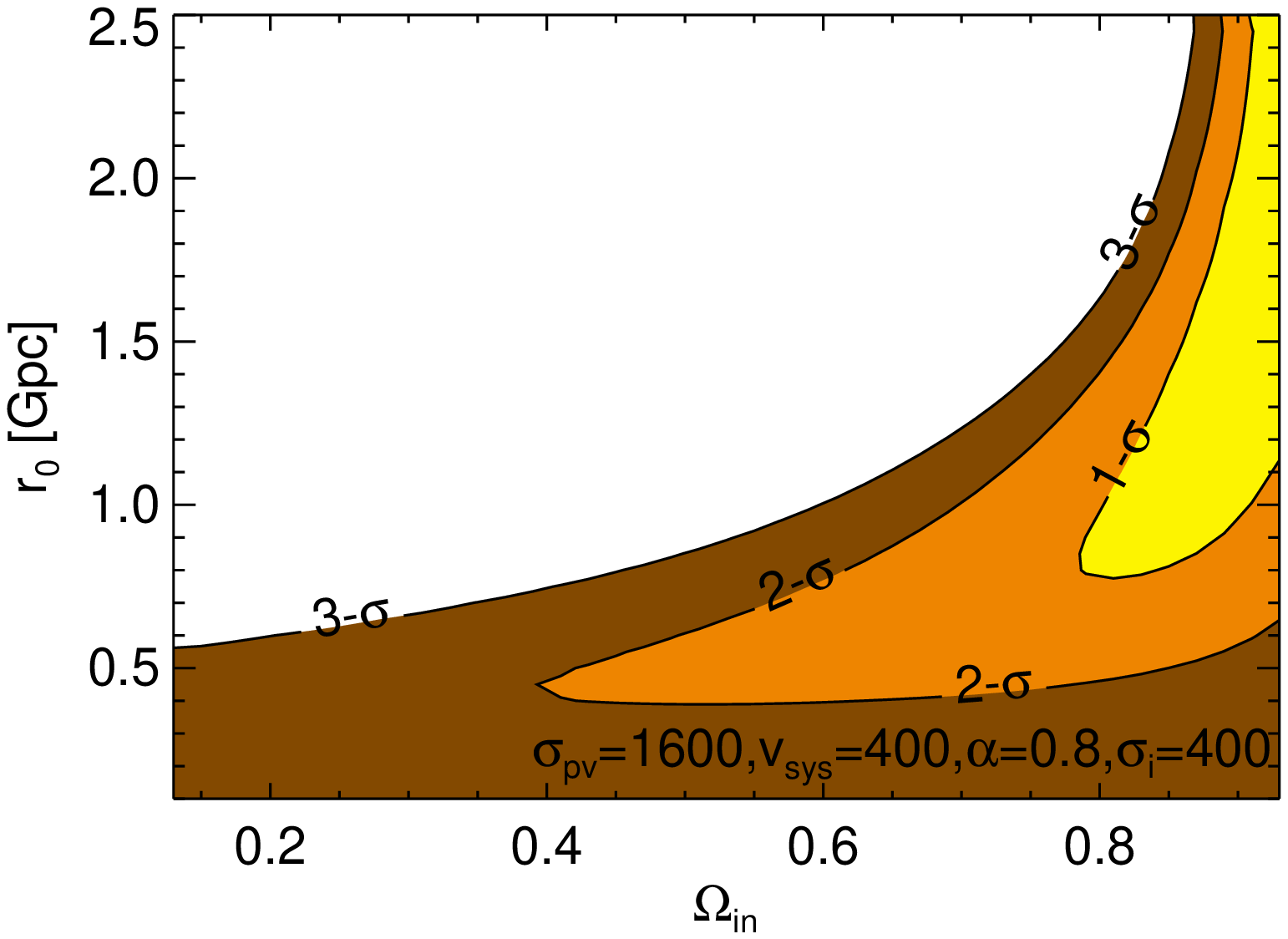} \\
\includegraphics[width=0.45 \textwidth]{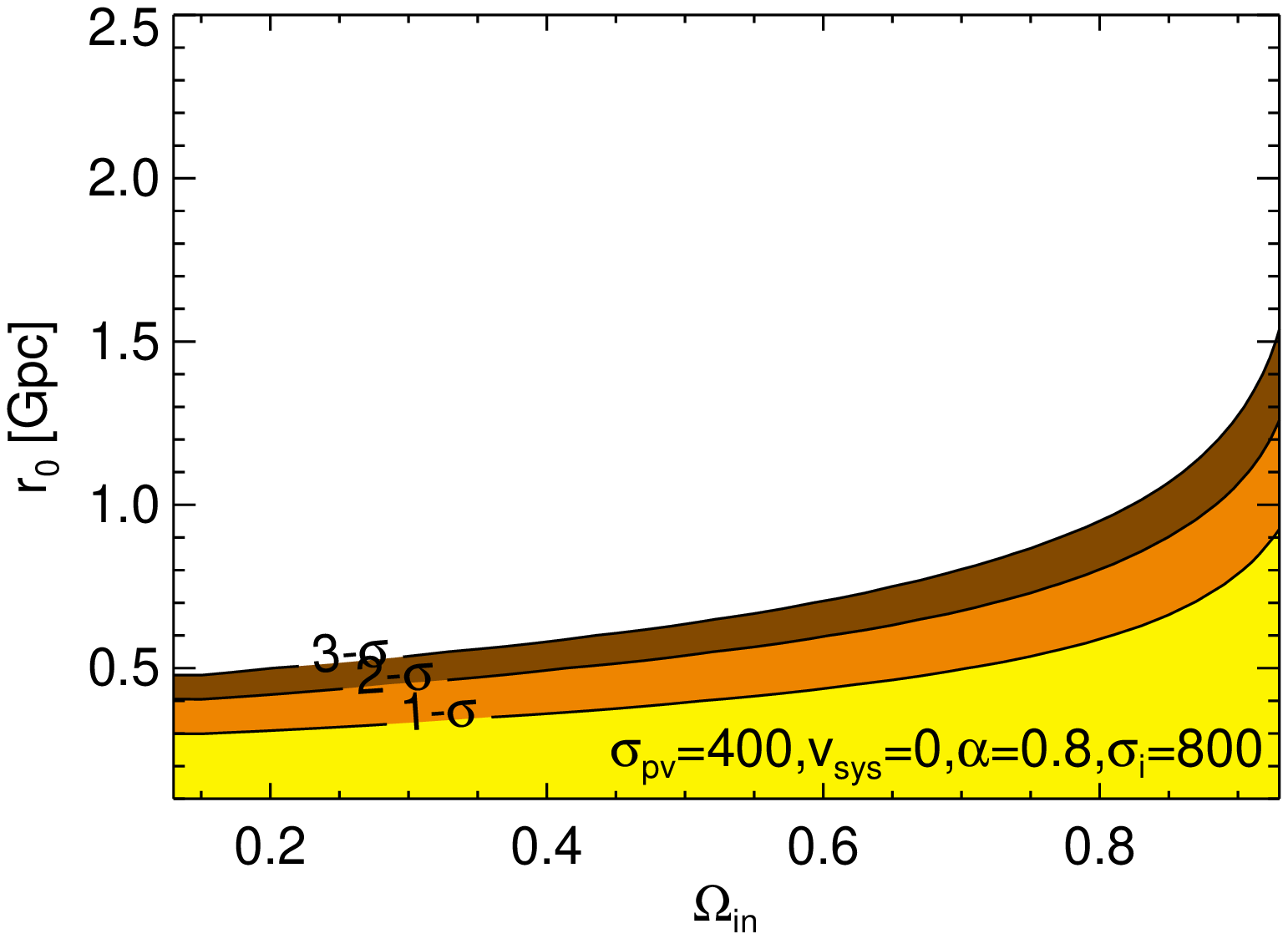}
\includegraphics[width=0.45 \textwidth]{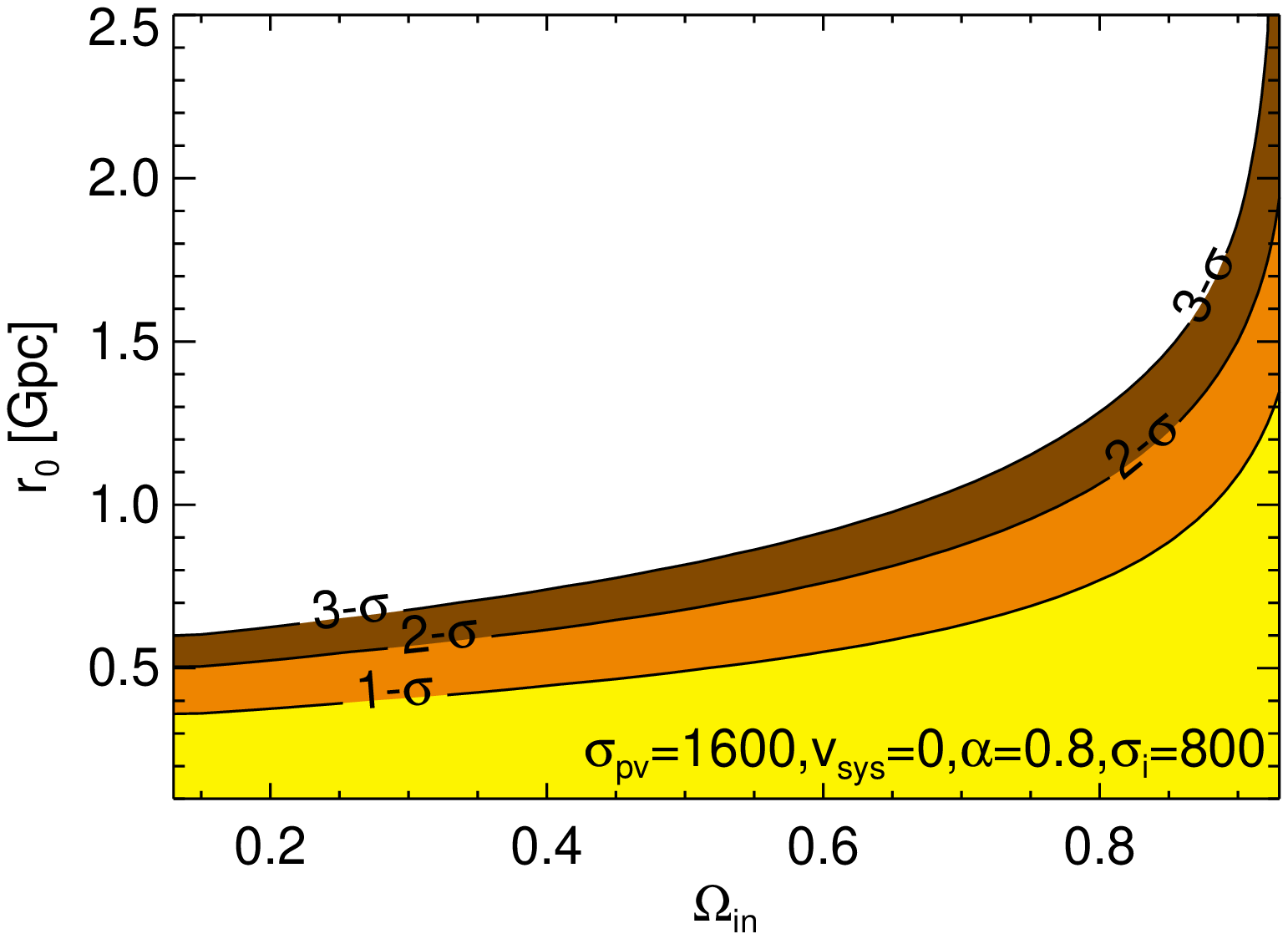} \\
\includegraphics[width=0.45 \textwidth]{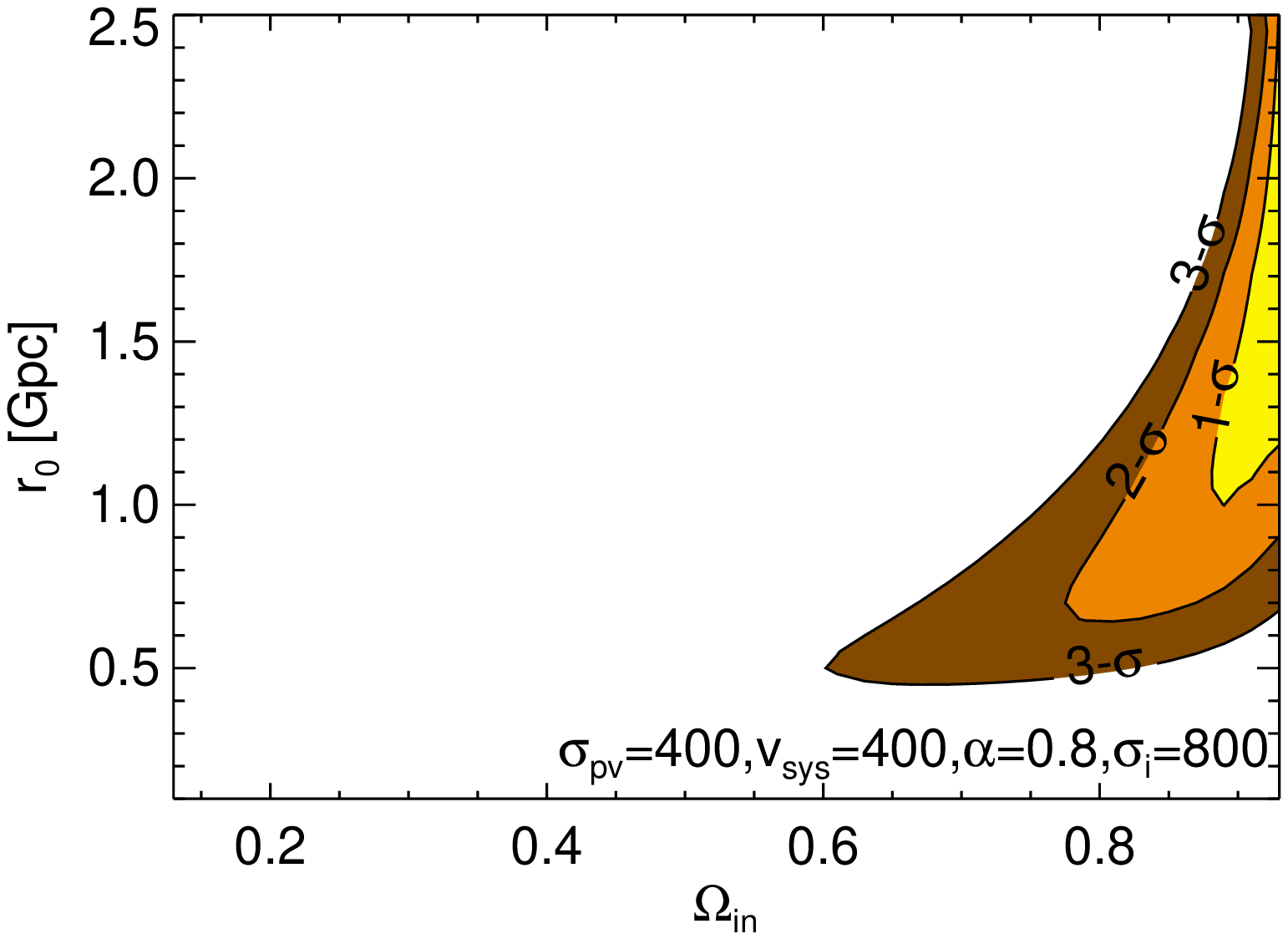}
\includegraphics[width=0.45 \textwidth]{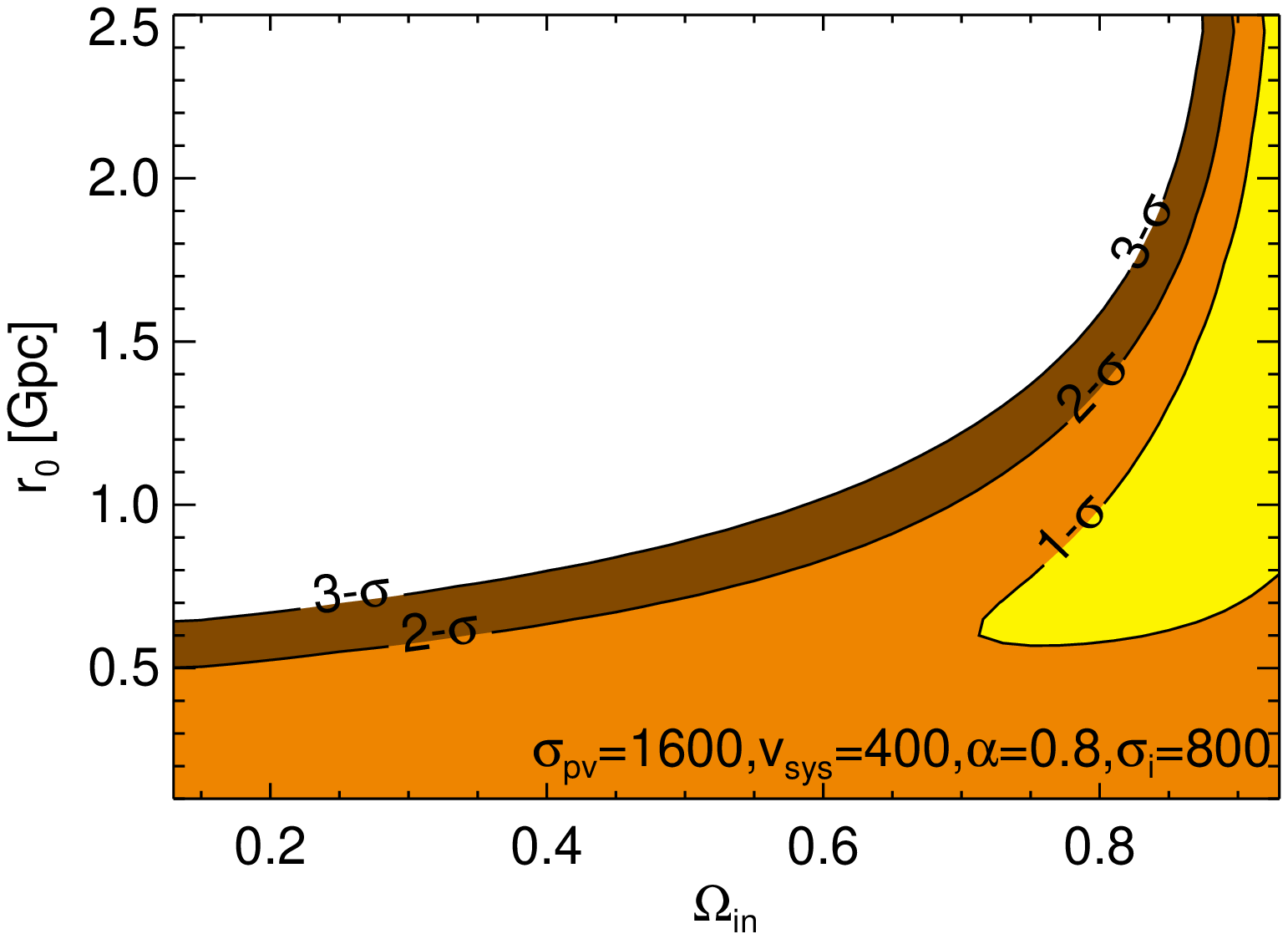}
\caption{Same as in Fig.~\ref{fig:forecastnc10} but for a
mock survey with 100 well observed clusters.}\label{fig:forecastnc100}
\end{center}
\end{figure}

We use Monte Carlo modelling to sample the PDF of the cluster redshift
distribution Eq.~\eqref{eq:nz}, and we assume that the clusters have
normal distributed peculiar velocities $v_i \sim
\mathcal{N}(0,\sigma_{\rm pv})$. In Figs.~\ref{fig:forecastnc10} and
\ref{fig:forecastnc100} are shown the 2D likelihoods in the cases were
we have 10 and 100 clusters with good kSZ measurements. As can be seen
already with 10 clusters, strong bounds can be put on the LTB model,
while with 100 clusters, assuming the FRW model is correct, giga
parsec sized voids are essentially ruled out.  Complementarily we have
shown that, if the measurements have systematic errors at the 400 km
s$^{-1}$ level, a spurious positive detection of a very shallow
underdensity ($\Omega_{\rm in} \lesssim 1$) could be inferred.
Conversely, care should be taken when constraining LTB models with kSZ
observations where systematic errors have been included
self-consistently in the analysis, under the assumption of a
homogeneous universe, by fixing the average peculiar velocity to zero,
as advocated in Ref.~\cite{Kosowsky:2008}.

\section{Discussion and conclusions}

As an alternative explanation of the dimming of distant supernovae it
has recently been advocated that we live in a special place in the
Universe near the centre of a large void. The universe is no longer
homogeneous and isotropic and the apparent late time acceleration is
actually a consequence of spatial gradients in the metric. If we did
not live close to the centre of the void, we would have observed a CMB
dipole much larger than that allowed by observations. Hence, until now
it has been argued, for the model to be consistent with observations,
that by coincidence we happen to live very close to the centre of the
void. However, even if we are at the centre of the void, we can
observe distant galaxy clusters, which are off-centre. In their frame
of reference there should be a large CMB dipole, which manifests
itself observationally for us as a kinematic Sunyaev-Zeldovich effect.

In this paper we have studied the induced dipole for off-centred
clusters due to the different trajectories of photons from the last
scattering surface, and computed the size of the corresponding
apparent velocity of those clusters with respect to the rest frame of
the CMB LSS, depending on the parameters of the void model. We then
analysed the present observations of the kSZ effect in a handful of
clusters and gave very strong constraints on the size of the void in
LTB models. In fact, for our specific constrained-GBH model the bounds
are in conflict with other constraints from supernovae, baryon
acoustic oscillations and CMB, at least at the 3-$\sigma$ level, see
Fig.~\ref{fig:otherdata}, and therefore we conclude that the
constrained-GBH void model is practically ruled out if the current
interpretation of kSZ observations are correct.

At present, kSZ data leave small voids ($r_0 \lesssim 800$ Mpc)
unconstrained, independently of the inner density contrast
($\Omega_{\rm in}$) used, simply because the radius is so small that
the void does not impact the cluster with the lowest redshift in the
sample, see Table~1. In order to put limits on small voids with large
density contrasts, or sudden transitions, we would need to use local
large scale structure data that can be checked for homogeneity, or
alternatively low redshift supernovae.

Current kSZ observations are limited in numbers, and are still not
precise enough to make a single positive detection of the peculiar
velocity of a cluster at the 2-$\sigma$ level, see
Fig.~\ref{fig:obs}.  Hence, one cannot rule out that they
are plagued by large systematic and/or random errors. This is about
to change in the coming years, and we have made predictions of how
systematic near-future observations from kSZ surveys like ACT or SPT
could strongly rule out all LTB models with giga parsec sized voids.
In the case that the LTB models were confirmed, the average apparent
velocity profile as a function of distance would give a direct
handle on the density and expansion rate of the void. This unique
relationship makes by far kSZ observations the most powerful data for
constraining LTB models.

If the sensitivity of experiments like ACT, SPT and Planck are as
planned, and the present systematic errors are under control, it is
expected that large voids will definitely be ruled out at many
sigma. If these experiments yield 10 (100) well observed clusters we
could reasonably expect to rule out voids of 800 (500) Mpc radius at
the 3-$\sigma$ level.  We hope to come back to this severe challenge
to LTB models and redo the analysis with the new data in the near
future.

\section*{Acknowledgments}

We thank Nick Kaiser for suggesting to us that kSZ data may severely
constrain LTB models, which prompted this investigation. We also thank
Nick and David Garfinkle for very interesting correspondence, and
Raul Jim\'enez for discussions on kSZ measurements with the ACT. 
We acknowledge the use of the computer resources of the Danish Centre
of Scientific Computing. We also acknowledge financial support from 
the Spanish Research Ministry, under contract FPA2006-05807 and
the Consolider Ingenio 2010 programme under contract CSD2007-00060
``Physics of the Accelerating Universe" (PAU).

\section*{References}
\bibliographystyle{hunsrt}
\bibliography{paper}

\end{document}